\documentclass[usegraphicx,usenatbib]{mn2e}
\usepackage{amsmath}
\usepackage{amssymb}
\usepackage{times}
\usepackage{subfigure}
\usepackage{epsfig}

\renewcommand{\descriptionlabel}[1]%
  {\hspace{\labelsep}\textbf{#1}}

\title[CCD time-series photometry of the globular cluster NGC 6981]
      {CCD time-series photometry of the globular cluster NGC 6981: Variable star census and physical parameter estimates}

\author[D.M. Bramich et al.]
  {D. M. Bramich$^{1}$\thanks{E-mail: dbramich@eso.org, dan.bramich@hotmail.co.uk},
   R. Figuera Jaimes$^{2}$\thanks{E-mail: rfiguera@ula.ve},
   Sunetra Giridhar$^{3}$\thanks{E-mail: giridhar@iiap.res.in},
   A. Arellano Ferro$^{2}$\thanks{E-mail: armando@astroscu.unam.mx}
  \medskip
  \\$^{1}$European Southern Observatory, Karl-Schwarzschild-Stra$\beta$e 2, 85748 Garching bei M\"{u}nchen, Germany
  \\$^{2}$Instituto de Astronom\'ia, Universidad Nacional Aut\'onoma de M\'exico, M\'exico
  \\$^{3}$Indian Institute of Astrophysics, Koramangala 560034, Bangalore, India
  }

\begin{document}

\date{Accepted 2010 August ???. Received 2010 August ???; Submitted 2010 August ???}

\pagerange{\pageref{firstpage}--\pageref{lastpage}} \pubyear{2010}

\maketitle

\label{firstpage}

\begin{abstract} 
We present the results from 10 nights of observations of the globular cluster NGC~6981 (M72) in the $V$, $R$ and $I$ Johnson wavebands.
We employed the technique of difference image analysis to perform precision differential photometry on the time-series
images, which enabled us to carry out a census of the under-studied variable star population of the cluster. We show that
20 suspected variables in the literature are actually non-variable, and we confirm the variable nature of another 29
variables while refining their ephemerides. We also detect 11 new RR Lyrae variables and 3 new SX Phe variables, bringing
the total confirmed variable star count in NGC~6981 to 43.

We performed Fourier decomposition of the light curves for a subset of RR Lyrae stars and used the Fourier parameters to estimate
the fundamental physical parameters of the stars using relations available in the literature. Mean values of these physical parameters
have allowed us to estimate the physical parameters of the parent cluster. We derive a metallicity of [Fe/H]$_{\mbox{\scriptsize ZW}} \approx -$1.48$\pm$0.03
on the \citet{zin1984} scale (or [Fe/H]$_{\mbox{\scriptsize UVES}} \approx -$1.38$\pm$0.03 on the new \citet{car2009} scale) for NGC~6981,
and distances of $\sim$16.73$\pm$0.36~kpc and $\sim$16.68$\pm$0.36~kpc from analysis of the RR0 and RR1 stars separately. We also confirm the Oosterhoff type
I classification for the cluster, and show that our colour-magnitude data is consistent with the age of $\sim$12.75$\pm$0.75~Gyr derived by \citet{dot2010}.
\end{abstract} 


\section{Introduction}
\label{sec:introduction}

The study of Galactic globular clusters is important for many reasons. These stellar systems represent some of the oldest, and consequently metal poor, stellar populations in the
Galaxy, and their scrutiny allows us to glean information regarding the formation and early evolution of the Galaxy. The spatial distribution of the clusters reveals
a different aspect of the Galactic structure than other stars in the Galaxy, and their orbits and tidal tails provide constraints on the Galactic potential. Of course,
what we learn about globular clusters in our own Galaxy is also applicable to other galaxies as well.

Globular clusters are also believed to be a close approximation to a stellar laboratory since a cluster's members were formed at the same time
from the same primordial material with the same composition, leading to a homogeneity of certain properties within each cluster, but with differences in these properties
between clusters. Although this paradigm is being challenged by the recent discovery in some globular clusters of multimodal main sequences and
sub-giant branches (\citealt{pio2009}, and references therein), indicating the existence of multiple stellar populations, most globular clusters
do not exhibit such obvious deviations from a simple stellar population and the paradigm still holds.

There are $\sim$150 globular clusters in our Galaxy for which their fundamental properties, such as metallicity, distance, age and
kinematics, have been estimated by various methods (\citealt{har1993}). One independent method for estimating at least some of these quantities
is by studying the population of RR Lyrae variable stars present in most clusters. This method uses the fact that the light curve morphology of
RR Lyrae stars is connected with their fundamental stellar parameters, and consequently quantities such as metallicity, absolute magnitude and effective
temperature may be calculated from the fit parameters of the Fourier decomposition of their light curves using empirical, semi-empirical or theoretical relations
published in recent years (\citealt{sim1993}; \citealt{jur1996}; \citealt{jur1998}; \citealt{kov1998}; \citealt{kov2001}; \citealt{mor2007}).
Appropriate mean values of these fundamental parameters then enable similar estimates of the physical parameters of the parent cluster.

As part of a series of papers on detecting and characterising the variable stars in globular clusters, and using the results to estimate
the parameters of the host cluster (\citealt{are2004}; \citealt{laz2006}; \citealt{are2008a}; \citealt{are2008b}; \citealt{are2010}), we
have performed CCD time-series photometry of the globular cluster NGC~6981 (RA~$\alpha = 20^{\mbox{\scriptsize h}} 53^{\mbox{\scriptsize m}} 27.9^{\mbox{\scriptsize s}}$,
Dec.~$\delta = -12\degr 32\arcmin 13\arcsec$, J2000; $l = 35.16\degr$, $b = -32.68\degr$) using the method of difference image analysis (Section~\ref{sec:observations_reductions}).
The known variables in this cluster, which are exclusively RR Lyrae variables, have been studied in a handful of photographic observing campaigns (\citealt{sha1920}; \citealt{ros1953};
\citealt{saw1953}; \citealt{nob1957}; \citealt{dic1972b}), the most recent of which is now 40 years in the past. Periods and ephemerides are poorly determined or non-existent
for a substantial number of variables, and light curves for many of the claimed ``variables'' have not been published. Therefore, in Section~\ref{sec:variable_stars}, we use
our precision differential photometry to perform an essential variable star census for the cluster.

In Section~\ref{sec:RRL_physical}, we use the Fourier decompostion of the RR Lyrae star light curves to estimate their fundamental physical parameters, and then in
Section~\ref{sec:cluster_properties}, we use the RR Lyrae properties to estimate the metallicity of, and distance to, NGC~6981. We also discuss the age estimates
for the cluster that are available in the literature in the context of our colour-magnitude diagram. Our conclusions are presented in Section~\ref{sec:conclusions}.

Throughout this paper we adopt the RR Lyrae nomenclature introduced by \citet{alc2000};
namely, RR0 designates an RR Lyrae star pulsating in the fundamental mode, and RR1 designates an RR Lyrae star pulsating in the first-overtone mode.

\section{Observations And Reductions}
\label{sec:observations_reductions}

\subsection{Observations}
\label{sec:observations}

We employed the 2.0m telescope of the Indian Astronomical Observatory (IAO), Hanle, India, located
at 4500m above sea level, to obtain time-series imaging of the globular cluster NGC~6981. The image data
were obtained during several runs between October 2004 and September 2009, where we collected a total
of 103, 110 and 3 images through Johnson $V$, $R$, $I$ filters, respectively
(see Table~\ref{tab:observations} for a detailed log of the observations).
The CCD camera that was used is equipped with a Thompson CCD of 2048$\times$2048~pixels with a pixel
scale of 0.296~arcsec~pix$^{-1}$ and a field-of-view of $\sim$10.1$\times$10.1~arcmin$^2$.

\begin{table}
\caption{The distribution of observations of NGC 6981 for each filter, where the columns $N_{V}$, $N_{R}$ and $N_{I}$ represent
         the number of images taken for the filters $V$, $R$ and $I$, respectively.
         We also provide the exposure time, or range of exposure times, employed during each night for each filter
         in the columns $t_{V}$, $t_{R}$ and $t_{I}$.}
\centering
\begin{tabular}{@{}lcccccc@{}}
\hline
Date     & $N_{V}$ & $t_{V}$ (s) & $N_{R}$ &  $t_{R}$ (s) & $N_{I}$ & $t_{I}$ (s) \\
\hline
20041004 & 20      &  60-180     & 19      & 100-150      & 0       & --- \\
20041005 & 29      &    90       & 29      &    70        & 0       & --- \\
20050514 & 12      &   150       & 15      &   120        & 0       & --- \\
20050515 & 13      &   150       & 15      &   120        & 0       & --- \\
20050516 &  5      & 100-150     &  6      &  80-120      & 0       & --- \\
20070522 &  4      & 200-300     &  5      & 200-240      & 0       & --- \\
20070804 &  5      &   240       &  8      &   180        & 0       & --- \\
20070905 & 11      &   150       & 11      &   120        & 0       & --- \\
20090913 &  4      &   150       &  1      &   120        & 3       & 100 \\
20090914 &  0      &   ---       &  1      &   120        & 0       & --- \\
\hline
Total:   & 103     &             & 110     &              & 3       &     \\
\hline
\end{tabular}
\label{tab:observations}
\end{table}

\subsection{Difference Image Analysis}
\label{sec:dia}

As in previous papers (\citealt{are2004}; \citealt{laz2006}; \citealt{are2008a}; \citealt{are2008b}; \citealt{are2010}),
we have employed the technique of difference image analysis (DIA) to extract high precision photometry
for all point sources in the images of NGC 6981, including those in the highly crowded central region
({\citealt{ala1998}; \citealt{ala2000}; \citealt{bra2005}). We used a pre-release version of the {\tt DanDIA}\footnote{
{\tt DanDIA} is built from the DanIDL library of IDL routines available at http://www.danidl.co.uk}
pipeline for the data reduction process (Bramich et al., in preparation) which includes a new algorithm
that models the convolution kernel matching the point-spread function (PSF) of a pair of images of the same field as a discrete pixel array
(\citealt{bra2008}).

In brief, the {\tt DanDIA} pipeline was used to perform the following processing steps on the raw image data. Bias level
and flat field corrections were applied to the raw images to create calibrated images, which were then 
cleaned of cosmic rays using the cosmic ray cleaning algorithm of \citet{van2001}. A reference image for each filter was
constructed by registering and stacking a set of the best-seeing calibrated images such that all images used were taken on a single night.
This resulted in 7, 7 and 2 images being stacked with total exposure times of 630, 490 and 200~s for the filters $V$, $R$ and $I$, respectively.
The full-width half-maximum (FWHM) of the PSF in the $V$, $R$ and $I$ filter reference images was measured to be
$\sim$3.8, $\sim$3.6 and $\sim$4.7~pix, respectively.

In each reference image, we measured the fluxes (referred to as reference fluxes) and positions of all PSF-like objects (stars)
by extracting a spatially variable (with polynomial degree 3) empirical PSF from the image and fitting this PSF to each detected object. Deblending of very close
objects was attempted. Using the Delaunay triangulation and triangle matching method of \citet{pal2006}, the detected stars in each image in the time-series
sequence were matched with those detected in the corresponding reference image, and a linear transformation was derived which was used to register each image with
the reference image using cubic O-MOMS resampling (\citealt{blu2001}).

For each filter, a sequence of difference images was created by subtracting the relevant reference image, convolved with an appropriate
spatially variable kernel, from each registered image. The spatially variable convolution kernel for each registered image was determined using bilinear interpolation of a
set of kernels, modelled as pixel arrays, that were derived for a uniform 6$\times$6 grid of subregions across the image.

\begin{figure}
\centering
\epsfig{file=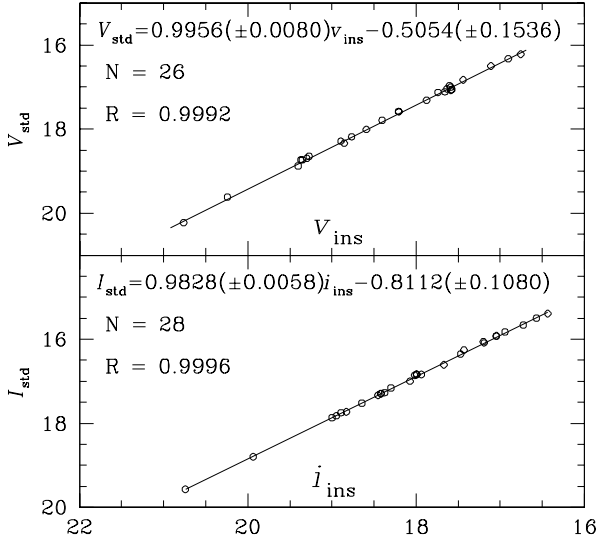,angle=0.0,width=\linewidth}
\caption{Plot of standard magnitude in the Johnson system against mean instrumental magnitude for the
         set of standard stars supplied by P.~Stetson that lie in our field of observation (open circles).
         The top and bottom panels correspond to the V and I filters, respectively. The solid line in   
         each panel shows the fitted relation between the standard and mean instrumental magnitudes.
         \label{fig:phot_calib}}
\end{figure}

The differential fluxes for each star detected in the reference image were measured on each difference image as follows. 
The empirical PSF at the measured position of the star on the reference image was determined by shifting the empirical PSF model
corresponding to the nearest pixel by the appropriate sub-pixel shift using cubic O-MOMS resampling. The empirical PSF was then convolved
with the kernel model corresponding to the star position and current difference image, and then optimally scaled to the difference image at the
star position using pixel variances $\sigma^{2}_{kij}$ for image $k$, pixel column $i$ and pixel row $j$, taken from the following
standard CCD noise model:
\begin{equation}
\sigma^{2}_{kij} = \frac{\sigma^{2}_{0}}{F^{2}_{ij}} + \frac{M_{kij}}{G F_{ij}}
\label{eqn:noise_model}
\end{equation}
where $\sigma_{0}$ is the CCD readout noise (ADU), $F_{ij}$ is the master flat-field image, $G$ is the CCD gain (e$^{-}$/ADU) and $M_{kij}$
is the image model (see \citealt{bra2008}).

Light curves for each star were constructed by calculating the total flux $f_{\mbox{\scriptsize tot}}(t)$ in ADU/s at each epoch $t$ from:
\begin{equation}
f_{\mbox{\scriptsize tot}}(t) = f_{\mbox{\scriptsize ref}} + \frac{f_{\mbox{\scriptsize diff}}(t)}{p(t)}
\label{eqn:totflux}
\end{equation}
where $f_{\mbox{\scriptsize ref}}$ is the reference flux (ADU/s), $f_{\mbox{\scriptsize diff}}(t)$ is the differential flux (ADU/s) and
$p(t)$ is the photometric scale factor (the integral of the kernel solution). Conversion to instrumental magnitudes was achieved using:
\begin{equation}
m_{\mbox{\scriptsize ins}}(t) = 25.0 - 2.5 \log (f_{\mbox{\scriptsize tot}}(t))
\label{eqn:mag}
\end{equation}
where $m_{\mbox{\scriptsize ins}}(t)$ is the instrumental magnitude of the star at time $t$. Uncertainties were propagated in the correct analytical fashion.

\subsection{Caveats Of Difference Imaging Analysis}
\label{sec:caveat}

\begin{figure*}
\centering
\begin{tabular}{cc}
\subfigure[]{\epsfig{file=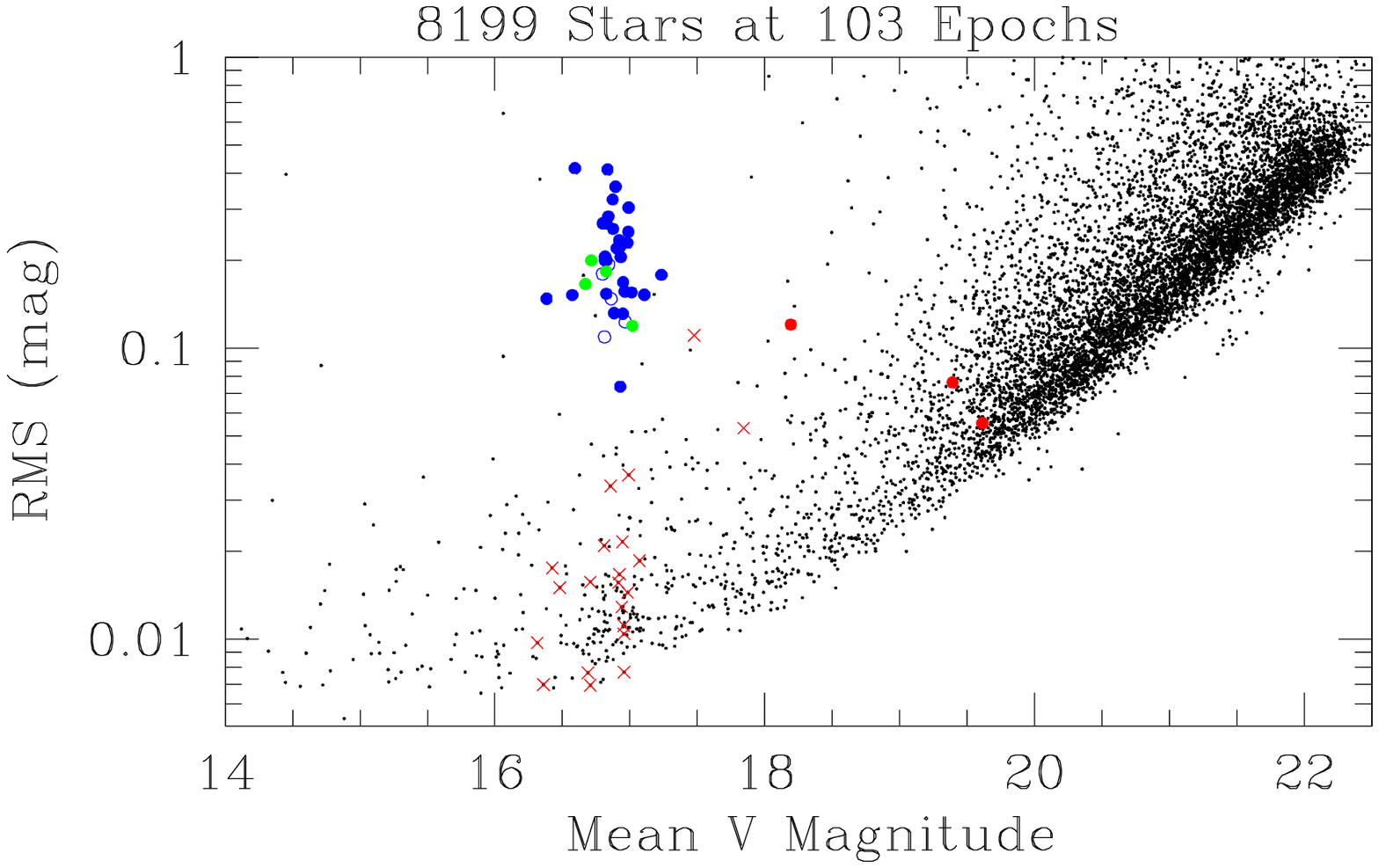,angle=0.0,width=0.48\linewidth} \label{fig:rms_diagram_V}} &
\subfigure[]{\epsfig{file=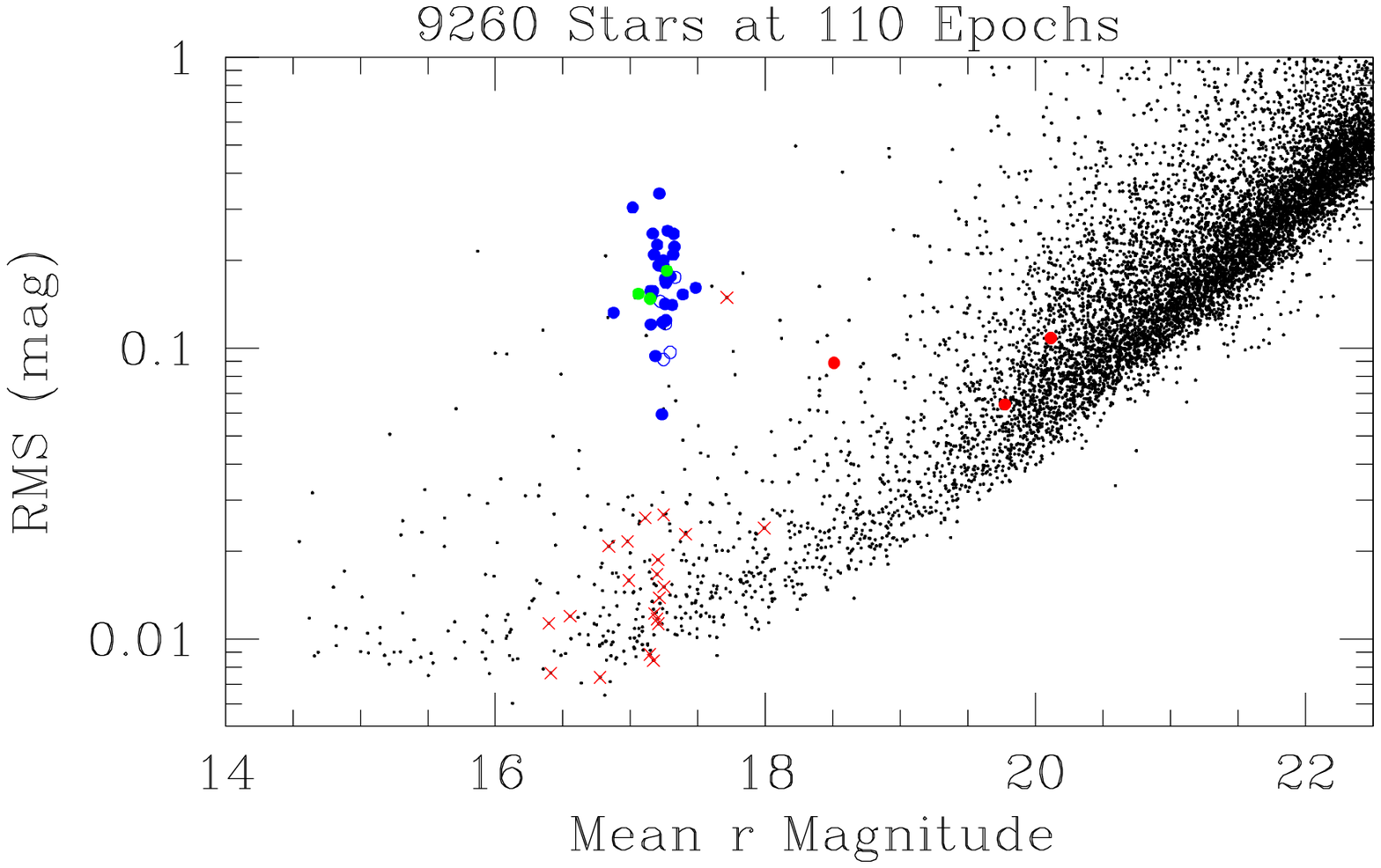,angle=0.0,width=0.48\linewidth} \label{fig:rms_diagram_r}} \\
\end{tabular}
\caption{(a) Plot of RMS magnitude deviation versus mean magnitude for each of the 8199 calibrated $V$ light curves.
         (b) Plot of RMS magnitude deviation versus mean magnitude for each of the 9260 instrumental $r$ light curves.
         Both plots: RR Lyrae variables of the RR0 and RR1 types are plotted as solid and open blue circles, respectively.
         Variables of the SX~Phe type are plotted as solid red circles, and suspected RR Lyrae variables are plotted as solid green circles.
         Previously suspected variables in the literature that do not show any variability in our data are marked as red crosses.
         A few of these previously suspected variables exhibit a relatively large RMS magnitude deviation which is due to outlier photometric
         measurements in the light curve rather than bona-fide variability.
         \label{fig:rms_diagrams}}
\end{figure*}

The value of the reference flux $f_{\mbox{\scriptsize ref}}$ for each star is measured on the reference image, and where the star field is very crowded, the measured $f_{\mbox{\scriptsize ref}}$ values
are likely to be systematically too large due to flux contamination from blending and unmodelled faint background stars (\citealt{tod2005}; \citealt{tod2006}). On the other hand,
the measurement of the difference fluxes $f_{\mbox{\scriptsize diff}}(t)$ does not suffer
from this problem because the majority of sources are fully subtracted in the difference images, including the blended stars and faint background stars. For a variable object, a
value for $f_{\mbox{\scriptsize ref}}$ that is systematically too large will result in a light curve with an amplitude in magnitudes that is systematically too small, and viceversa 
(Equations~\ref{eqn:totflux}~\&~\ref{eqn:mag}), although the ``shape'' of the light curve will be unaffected.
Consequently, we have made a note in column 4 of Table~\ref{tab:variables} (see later) of those variable stars likely to be affected by
flux contamination, or that are affected by a blended PSF.

Furthermore, it is worth mentioning a feature of the difference image construction with {\tt DanDIA} which can impede photometry on some objects.
Saturated pixels in the reference image are flagged as bad pixels, and when the reference image is convolved with a kernel in order to match the PSF of the current registered image,
the bad pixels in the reference image are grown by the footprint of the convolution kernel. For our data, we modelled the kernel as a circular pixel array
of radius equal to twice the PSF FWHM in the current registered image. Hence a saturated star in the reference image discounts an area in the difference images that 
encompasses the saturated star and its immediate neighbourhood, with poorer seeing resulting in a larger discounted area. Consequently stars in the neighbourhood of saturated stars
may suffer from imprecise photometry (because fewer good pixels are available in the difference images for photometric measurements), missing photometric measurements for a 
subset of epochs
(when the seeing is too poor), or simply it may have been impossible to extract any photometric measurements at all for the star in question.

We made our choice of images to be combined into the stacked reference images so as to minimise the number of saturated stars.
However, our reference images still contain a handful of saturated stars towards the centre of the cluster, which has affected the photometry of some of the variable stars.
We will refer to this point a number of times later in this paper.

\subsection{Photometric Calibrations}
\label{sec:photcal}

We derived photometric calibration relations for the conversion of instrumental $v$ and $i$ magnitudes to the Johnson-Kron-Cousins
photometric system (\citealt{lan1992}) by analysing the standard $V$ and $I$ magnitudes of a set of standard stars in the
observed field of the cluster which were kindly made
available to us by P.~Stetson (private communication). We identified 26 and 28 standard stars in the $V$ and $I$ reference images,
respectively, and fitted a linear relation between standard and mean instrumental magnitudes for each filter (see Figure~\ref{fig:phot_calib}).
No colour term was found to be significant, and the linear correlation coefficients are $>$0.999. These relations were then
used to convert the instrumental $v$ and $i$ photometric measurements for all detected point sources to the standard Johnson-Kron-Cousins photometric
system.

In Figure~\ref{fig:rms_diagram_V} we plot the root-mean-square (RMS) magnitude deviation for each of the 8199 calibrated $V$ light curves versus the mean magnitude.
We achieve better than 20~mmag scatter for stars in the magnitude range 14 to $\sim$18.5~mag, and $\sim$6-10~mmag photometry at the bright end.

The instrumental $r$ magnitudes were retained in the instrumental system since no standards with $R$ magnitude measurements
were found in the literature. In Figure~\ref{fig:rms_diagram_r} we plot the RMS magnitude deviation for each of the 9260 instrumental $r$ light curves versus the mean magnitude.
In this filter we achieve $\sim$8-12~mmag photometry at the bright end.

\subsection{Astrometry}
\label{sec:astrometry}

A linear astrometric solution was derived for the $V$ filter reference image by matching $\sim$1000 hand-picked stars with the USNO-B1.0 star catalogue
(\citealt{mon2003}) using a field overlay in the image display tool {\tt GAIA} (\citealt{dra2000}). We achieved a radial RMS scatter in the residuals of
$\sim$0.3~arcsec. The astrometric fit was then used to calculate the J2000.0 celestial coordinates for all the confirmed variables in our field of view
(see Table~\ref{tab:astrom}). The coordinates correspond to the epoch of the $V$ reference image, which is the heliocentric Julian date $\sim$2453284.11~d.

\section{Variable Stars In NGC 6981}
\label{sec:variable_stars}

\begin{table*}
\caption{Details of all confirmed variables in NGC~6981. The variable star ``V'' identification numbers are listed in the first column, and the previous star identification
         numbers (for the \citet{kad1995} candidate variables) are listed in the second column. The third column indicates the variable star type.
         A guide to the level of blending of the variable star PSF in the reference images is provided in column 4,
         where ``Inner'' and ``Outer'' indicate that the variable is inside or outside, respectively, of the highly crowded central area of the cluster (the dividing
         line is placed at $r=\,$50~arcsec), and where ``Blend'' indicates that the variable is blended with a star of similar or greater brightness.
         In column 5, we list the number $N_{\mbox{\scriptsize max}}$ and $N_{\mbox{\scriptsize min}}$ of light curve maxima and minima, respectively,
         that were observed in full during our observation runs. Note that the epoch of maximum light $T_{\mbox{\scriptsize max}}$ in column 6 is a heliocentric Julian date (HJD).
         To enable easy referencing, the period for each variable as found by previous authors is supplied in columns 8 to 11, and may be compared to our derived
         periods in column 7. Note that most of the period estimates from previous authors fail to phase our light curves properly, which highlights the importance
         of the longer temporal baseline and much better photometric precision of our data.
         }
\centering   
\begin{tabular}{@{}lllllllllll@{}}
\hline
         &           &               &       & No. Observed                                               &                              &                 & Period        & Period          & Period          & Period From      \\
         & Candidate &               &       & Maxima \&                                                  &                              &                 & From          & From            & From            & Dickens \&       \\
Variable & Variable  & Variable      & Blend & Minima                                                     & $T_{\mbox{\scriptsize max}}$ & P               & SH1920        & Rosino          & Nobili          & Flinn (1972)     \\
Star ID  & Star ID   & Type          & Guide & $(N_{\mbox{\scriptsize max}}, N_{\mbox{\scriptsize min}})$ & (d)                          & (d)             & (d)           & (1953) (d)      & (1957) (d)      & (d)              \\
\hline
V1       & ---       & RR0           & Outer & (2,0)                                                      & 2453505.396                  & 0.619782        & 0.61974$^{h}$ & 0.619818$^{h}$  & ---             & ---              \\
V2       & ---       & RR0           & Outer & (0,1)                                                      & 2454317.174$^{d}$            & 0.465254        & 0.46561$^{h}$ & 0.4652687$^{h}$ & ---             & 0.46526213$^{h}$ \\
V3       & ---       & RR0           & Outer & (1,3)                                                      & 2454317.333                  & 0.497614        & 0.48965$^{h}$ & 0.4976104       & ---             & 0.4976052        \\
V4       & ---       & RR0           & Outer & (1,1)                                                      & 2454349.158                  & 0.552486        & 0.3619$^{h}$  & 0.5524877       & ---             & 0.5524863        \\
V5       & ---       & RR0           & Inner & (2,3)                                                      & 2453505.421                  & 0.507264        & 0.4991$^{h}$  & ---             & ---             & ---              \\
V7       & ---       & RR0           & Inner & (1,2)                                                      & 2453283.202                  & 0.524686        & 0.52463$^{h}$ & ---             & 0.524648$^{h}$  & 0.524630$^{h}$   \\
V8       & ---       & RR0           & Outer & (1,1)                                                      & 2453284.107                  & 0.568380        & 0.5743$^{h}$  & 0.568392$^{h}$  & ---             & 0.5683752$^{h}$  \\
V9       & ---       & RR0           & Inner & (2,0)                                                      & 2453505.421                  & 0.602928        & 0.5902$^{h}$  & ---             & ---             & 0.60296$^{h}$    \\
V10      & ---       & RR0           & Outer & (0,1)                                                      & 2454243.386$^{d}$            & 0.558186        & 0.5483$^{h}$  & 0.5581805       & ---             & 0.5581814        \\
V11      & ---       & RR0           & Outer & (0,2)                                                      & 2454317.325$^{d}$            & 0.520676        & 0.3345$^{h}$  & 0.521466$^{h}$  & ---             & 0.51997$^{h}$    \\
V12      & ---       & RR1           & Inner & (2,1)                                                      & 2453506.374                  & 0.287858        & 0.4111$^{h}$  & ---             & ---             & ---              \\
V13      & ---       & RR0           & Inner & (2,2)                                                      & 2453283.190                  & 0.542035        & 0.54182$^{h}$ & ---             & ---             & 0.55114$^{h}$    \\
V14      & ---       & RR0           & Inner & (0,1)                                                      & 2453507.413$^{d}$            & 0.607194        & 0.5904$^{h}$  & ---             & ---             & ---              \\
V15      & ---       & RR0           & Outer & (0,0)                                                      & 2454243.370$^{d}$            & 0.540460        & 0.5499$^{h}$  & 0.5403524$^{h}$ & ---             & 0.55044$^{h}$    \\
V16      & ---       & RR0           & Inner & (1,2)                                                      & 2454349.176                  & 0.575211        & 0.5641$^{h}$  & ---             & ---             & 0.585497$^{h}$   \\
V17      & ---       & RR0           & Inner & (1,1)                                                      & 2453283.140                  & 0.573540        & 0.56308$^{h}$ & 0.573539        & ---             & 0.5735404        \\
V18      & ---       & RR0           & Blend & (0,0)                                                      & 2454317.201$^{d}$            & 0.535578        & 0.52016$^{h}$ & ---             & ---             & ---              \\
V20      & ---       & RR0           & Inner & (0,1)                                                      & 2453506.453$^{d}$            & 0.595048        & 0.59555$^{h}$ & 0.595046        & ---             & ---              \\
V21      & ---       & RR0           & Outer & (1,0)                                                      & 2453506.410                  & 0.531162        & 0.5310$^{h}$  & 0.5311618       & ---             & 0.5311636        \\
V23      & ---       & RR0           & Outer & (0,0)                                                      & 2454317.325$^{d}$            & 0.585127        & 0.5969$^{h}$  & 0.5850834$^{h}$ & ---             & 0.585083$^{h}$   \\
V24      & ---       & RR1           & Blend & (2,2)                                                      & 2453505.448                  & 0.327129        & 0.4973$^{h}$  & ---             & ---             & ---              \\
V25      & ---       & RR1           & Outer & (0,3)                                                      & 2455088.167$^{d}$            & 0.353340        & ---           & 0.3533494$^{h}$ & ---             & 0.3533739$^{h}$  \\
V27      & ---       & RR0           & Outer & ---                                                        & ---                          & ---             & 0.65885       & 0.6739040       & ---             & 0.673774         \\
V28      & ---       & RR0           & Outer & (1,0)                                                      & 2454243.270                  & 0.567216        & 0.36381$^{h}$ & 0.5672533$^{h}$ & ---             & 0.56724873$^{h}$ \\
V29      & ---       & RR0           & Outer & (0,0)                                                      & 2453507.446$^{d}$            & 0.597472$^{e}$  & 0.36865$^{h}$ & ---             & 0.373614$^{h}$  & 0.605497$^{h}$   \\
V31      & ---       & RR0           & Inner & (1,1)                                                      & 2453283.181                  & 0.542326        & 0.55465$^{h}$ & ---             & ---             & 0.53249$^{h}$    \\
V32      & ---       & RR0           & Outer & (0,0)                                                      & 2454317.333$^{d}$            & 0.528299        & 0.50544$^{h}$ & 0.5282821$^{h}$ & ---             & 0.52834$^{h}$    \\
V35      & ---       & RR0           & Outer & ---                                                        & ---                          & ---             & ---           & ---             & 0.54374         & 0.543771         \\
V36      & ---       & RR0           & Blend & (0,1)                                                      & 2454243.450$^{d}$            & 0.582613        & ---           & ---             & ---             & ---              \\
V43      & S3        & RR1           & Inner & (2,1)                                                      & 2453283.195                  & 0.283493        & ---           & ---             & ---             & ---              \\
V44      & S5        & RR0$^{a}$     & Inner & (0,0)                                                      & ---                          & ---             & ---           & ---             & ---             & ---              \\
V45      & S7        & RR0$^{a}$     & Blend & (0,0)                                                      & ---                          & ---             & ---           & ---             & ---             & ---              \\
V46      & S8        & RR1           & Inner & (2,1)                                                      & 2453284.190                  & 0.286685        & ---           & ---             & ---             & ---              \\
V47      & S9        & RR0           & Inner & (1,0)                                                      & 2453284.221                  & 0.649084        & ---           & ---             & ---             & ---              \\
V48      & R3        & RR0           & Inner & (1,0)                                                      & 2453284.066                  & 0.639764        & ---           & ---             & ---             & ---              \\
V49      & R4        & RR0           & Inner & (0,1)                                                      & 2454243.387$^{d}$            & 0.578270        & ---           & ---             & ---             & ---              \\
V50      & ---       & RR0           & Inner & (1,2)                                                      & 2453284.244                  & 0.488880        & ---           & ---             & ---             & ---              \\
V51      & ---       & RR0$^{b}$     & Blend & (0,0)                                                      & 2454243.450$^{d}$            & 0.357335$^{f}$  & ---           & ---             & ---             & ---              \\
V52      & ---       & RR0/RR1$^{c}$ & Blend & (1,0)                                                      & 2453284.137                  & ---             & ---           & ---             & ---             & ---              \\
V53      & ---       & RR0/RR1$^{c}$ & Inner & (1,0)                                                      & 2453284.215                  & ---             & ---           & ---             & ---             & ---              \\
V54      & ---       & SX Phe        & Inner & (6,4)                                                      & 2453283.162                  & 0.0719862$^{g}$ & ---           & ---             & ---             & ---              \\
V55      & ---       & SX Phe        & Outer & (11,11)                                                    & 2453283.132                  & 0.0470327$^{g}$ & ---           & ---             & ---             & ---              \\
V56      & ---       & SX Phe        & Inner & (9,9)                                                      & 2453283.158                  & 0.0404696$^{g}$ & ---           & ---             & ---             & ---              \\
\hline
\end{tabular}
\raggedright
$^{a}$These variables are RR Lyrae stars, and we believe that they are of the RR0 type, but we cannot confirm this (see Section~\ref{sec:unknown}).
$^{b}$The classification for this variable is most likely RR0 (see Section~\ref{sec:RRL_stars}).
$^{c}$These variables are RR Lyrae stars, but we have been unable to distinguish their sub-type (see Section~\ref{sec:unknown}).
$^{d}$The epoch of maximum light is uncertain because we have not observed the light curve peak. The epoch reported here is that of the data point closest to the
suspected peak.
$^{e}$The period listed for V29 is the period $P_{0}$ in Equation~\ref{eqn:per_change}. We detect a secular period change for this star of
$\beta\approx-$1.38$\times$10$^{-8}$~d~d$^{-1}$ (see Section~\ref{sec:RRL_stars}).
$^{f}$Due to the poor phase coverage of our observations for this variable, we have been unable to determine a reliable period (see Section~\ref{sec:RRL_stars}).
$^{g}$The periods listed for the SX~Phe variables are the periods associated with the largest amplitude oscillation.
$^{h}$Our photometric data are not well-phased by these periods.
\label{tab:variables}
\end{table*}

\begin{figure*} 
\centering
\epsfig{file=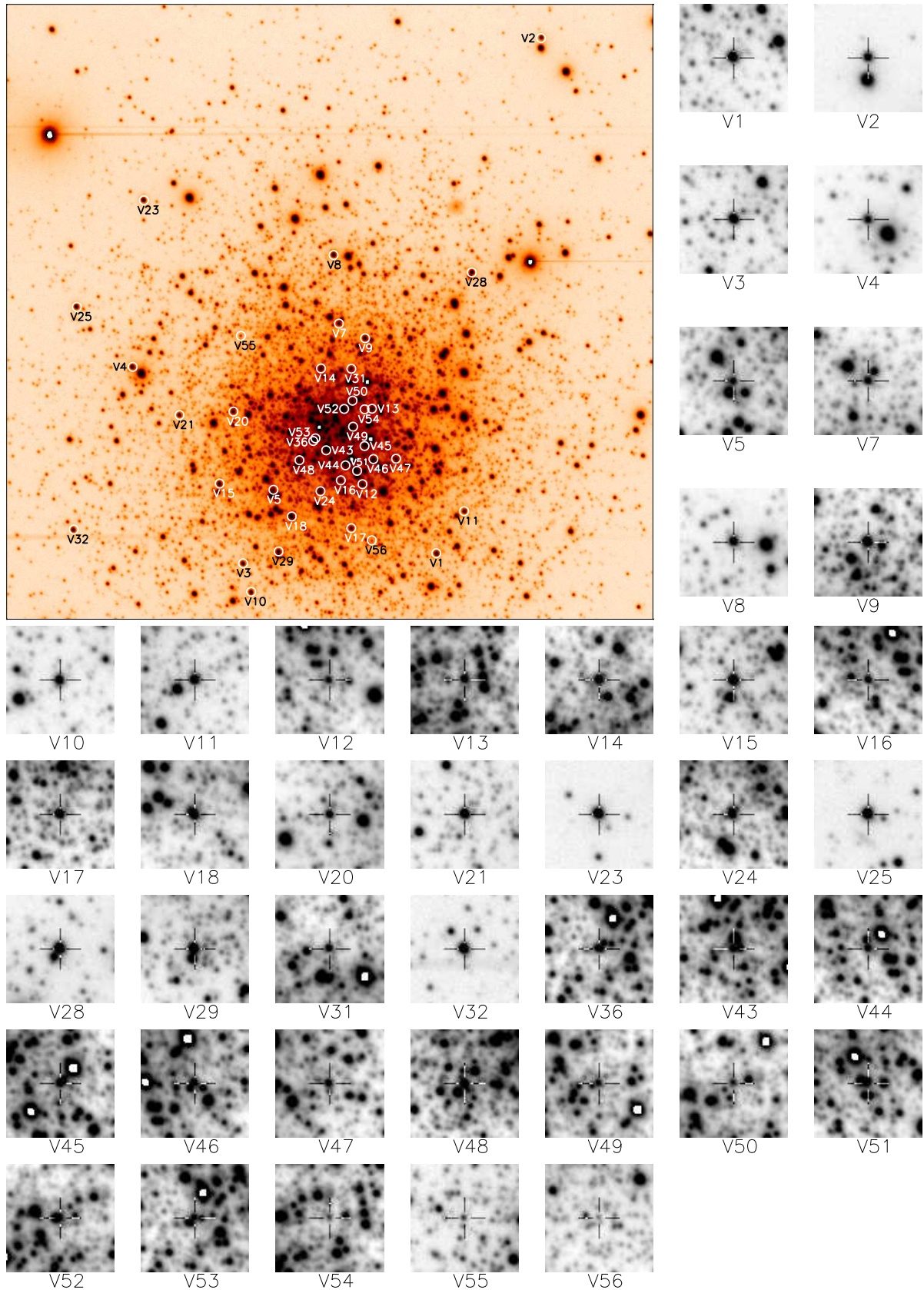,angle=0.0,width=0.9\linewidth}
\caption{Finding charts constructed from our $V$ reference image; north is up and east is to the right. The cluster image is 5.31$\arcmin$ by 4.93$\arcmin$, and the image stamps are
         of size 23.7$\arcsec$ by 23.7$\arcsec$. Each confirmed variable lies at the centre of its corresponding image stamp and
         is marked by a cross-hair. Note that finding charts for V27 and V35 are not available from our data because
         these variables are not in our field of view. The best finding chart for these stars may be found in \citet{dic1972a}.
         \label{fig:finding_charts}}
\end{figure*}

\begin{table*}
\caption{Celestial coordinates for all the confirmed variables in our field of view, except V27 and V35 which lie outside of our field of view.
         The coordinates correspond to the epoch of the $V$ reference image, which is the heliocentric Julian date $\sim$2453284.11~d.
         }
\centering
\begin{tabular}{ccccccccc}
\hline
Variable     & RA          & Dec.        & Variable     & RA          & Dec.        & Variable     & RA          & Dec.        \\
Star ID      & (J2000.0)   & (J2000.0)   & Star ID      & (J2000.0)   & (J2000.0)   & Star ID      & (J2000.0)   & (J2000.0)   \\
\hline
V1           & 20 53 31.12 & -12 33 11.9 & V16          & 20 53 27.86 & -12 32 36.9 & V44          & 20 53 28.02 & -12 32 29.7 \\
V2           & 20 53 34.56 & -12 29 02.3 & V17          & 20 53 28.23 & -12 33 00.0 & V45          & 20 53 28.66 & -12 32 20.2 \\
V3           & 20 53 24.57 & -12 33 17.3 & V18          & 20 53 26.21 & -12 32 54.6 & V46          & 20 53 28.97 & -12 32 26.5 \\
V4           & 20 53 20.79 & -12 31 42.6 & V20          & 20 53 24.21 & -12 32 04.0 & V47          & 20 53 29.73 & -12 32 26.3 \\
V5           & 20 53 25.58 & -12 32 41.7 & V21          & 20 53 22.38 & -12 32 05.8 & V48          & 20 53 26.46 & -12 32 27.3 \\
V7           & 20 53 27.77 & -12 31 21.1 & V23          & 20 53 21.13 & -12 30 21.9 & V49          & 20 53 28.27 & -12 32 10.9 \\
V8           & 20 53 27.57 & -12 30 47.9 & V24          & 20 53 27.18 & -12 32 42.2 & V50          & 20 53 28.25 & -12 31 58.3 \\
V9           & 20 53 28.66 & -12 31 28.0 & V25          & 20 53 18.88 & -12 31 13.7 & V51          & 20 53 28.41 & -12 32 32.3 \\
V10          & 20 53 24.85 & -12 33 31.1 & V28          & 20 53 32.26 & -12 30 55.9 & V52          & 20 53 27.96 & -12 32 02.3 \\
V11          & 20 53 32.05 & -12 32 51.5 & V29          & 20 53 25.77 & -12 33 11.5 & V53          & 20 53 27.00 & -12 32 16.7 \\
V12          & 20 53 28.60 & -12 32 38.7 & V31          & 20 53 28.20 & -12 31 42.9 & V54          & 20 53 28.66 & -12 32 02.5 \\
V13          & 20 53 28.91 & -12 32 02.2 & V32          & 20 53 18.82 & -12 33 01.5 & V55          & 20 53 24.45 & -12 31 27.3 \\
V14          & 20 53 27.17 & -12 31 42.8 & V36          & 20 53 26.92 & -12 32 18.0 & V56          & 20 53 28.93 & -12 33 05.9 \\
V15          & 20 53 23.76 & -12 32 38.8 & V43          & 20 53 27.36 & -12 32 22.4 &              &             &             \\
\hline
\end{tabular}
\label{tab:astrom}
\end{table*}

\begin{table*}
\caption{Time-series $V$, $r$ and $I$ photometry for all the confirmed variables in our field of view, except V27 and V35 which lie outside of our field of view.
         The standard $M_{\mbox{\scriptsize std}}$ and instrumental $m_{\mbox{\scriptsize ins}}$ magnitudes are listed in columns 4 and 5, respectively, corresponding to the variable
         star, filter, and epoch of mid-exposure listed in columns 1-3, respectively (note that we do not supply standard $R$ magnitudes). The uncertainty on $m_{\mbox{\scriptsize ins}}$ is listed
         in column 6, which also corresponds to the uncertainty on $M_{\mbox{\scriptsize std}}$. For completeness, we also list the quantities $f_{\mbox{\scriptsize ref}}$, $f_{\mbox{\scriptsize diff}}$
         and $p$ from Equation~\ref{eqn:totflux} in columns 7, 9 and 11, along with the uncertainties $\sigma_{\mbox{\scriptsize ref}}$ and $\sigma_{\mbox{\scriptsize diff}}$
         in columns 8 and 10.
         This is an extract from the full table, which is available with the electronic version of the article (see Supporting Information).
         }
\centering
\begin{tabular}{ccccccccccc}
\hline
Variable & Filter & HJD & $M_{\mbox{\scriptsize std}}$ & $m_{\mbox{\scriptsize ins}}$ & $\sigma_{m}$ & $f_{\mbox{\scriptsize ref}}$ & $\sigma_{\mbox{\scriptsize ref}}$ & $f_{\mbox{\scriptsize diff}}$ &
$\sigma_{\mbox{\scriptsize diff}}$ & $p$ \\
Star ID  &        & (d) & (mag)                        & (mag)                        & (mag)        & (ADU s$^{-1}$)               & (ADU s$^{-1}$)                    & (ADU s$^{-1}$)                &
(ADU s$^{-1}$)                     &     \\
\hline
V1 & $V$ & 2453283.09908    & 16.954 & 17.537 & 0.006  & 1242.465 & 0.924  & -274.361 & 5.059  & 0.9944 \\
V1 & $V$ & 2453283.10500    & 16.949 & 17.531 & 0.003  & 1242.465 & 0.924  & -270.162 & 2.872  & 0.9982 \\
\vdots   & \vdots & \vdots  & \vdots & \vdots & \vdots & \vdots   & \vdots & \vdots   & \vdots & \vdots \\
V1 & $R$ & 2453283.10977    & 0.000  & 17.241 & 0.003  & 1520.098 & 1.141  & -248.860 & 3.744  & 0.9937 \\ 
V1 & $R$ & 2453283.11633    & 0.000  & 17.256 & 0.004  & 1520.098 & 1.141  & -266.380 & 4.027  & 0.9950 \\
\vdots   & \vdots & \vdots  & \vdots & \vdots & \vdots & \vdots   & \vdots & \vdots   & \vdots & \vdots \\
V1 & $I$ & 2455088.17344    & 16.544 & 17.659 & 0.005  & 862.748  & 4.245  & 0.747    & 4.130  & 0.9995 \\   
V1 & $I$ & 2455088.17589    & 16.547 & 17.662 & 0.005  & 862.748  & 4.245  & -1.156   & 4.252  & 1.0009 \\
\vdots   & \vdots & \vdots  & \vdots & \vdots & \vdots & \vdots   & \vdots & \vdots   & \vdots & \vdots \\
\hline
\end{tabular}
\label{tab:vri_phot}
\end{table*}

The first claimed detection of two variable stars in NGC~6981 was made by \citet{dav1917}.
However, the first proper study of the variable stars in NGC 6981 was undertaken by \citet{sha1920} (from now on SH1920) where 18 photographic plates
taken over 2 years from the 60-inch reflector at the Mount Wilson Observatory were analysed. Stars V1-V34 were listed as ``probably
variable'' by Miss Ritchie and photometric measurements were extracted relative to a set of 29 comparison stars
of ``sensibly constant light''. However, it was noted that stars V6, V19, V22, V26 and V33 ``do not appear to be
conspicuously variable'', and that V25, V30 and V34 ``undoubtedly vary, but it has not been possible to obtain
uniform periods for them''. Hence SH1920 actually claim detection of 29 variables, and provide periods
for 26 of them.

\citet{ros1953} studied 22 of the claimed variables from the work of SH1920 and estimated new periods for 16
of them. Around the same time, \citet{saw1953} claimed the detection of 7 more variables (stars V35-V41) from a collection of 61 photographic plates,
although no light curves were published and no periods were estimated. \cite{nob1957} further studied the stars V7, V29 and V35, presenting light curves
and new periods. 

Later on, \citet{dic1972b} analysed $B$ and $V$ filter observational data for 21 RR Lyrae stars with
clear variations, using $\sim$20 photographic plates for each filter taken on 13 nights spread over the period of $\sim$1~year.
They revise the periods of these stars by combination with previous data, and attempt to use the RR Lyraes to derive various cluster
properties. From the same data, \citet{dic1973} claim the discovery of ``a new red variable'', but no light curve or period was ever
published. This star was subsequently labelled as V42 in the \citet{saw1973} catalogue.

In a brief paper by \citet{kad1995}, the authors use the cluster colour magnitude diagram to identify 16 stars in the RR Lyrae instability
strip that are not already claimed to be variable in previous publications. The stars, labelled S1-S9 and R1-R7, are put forwards as
``suspected'' and ``possible'' variables, respectively. Although these stars are not confirmed variables, they would be good candidates 
to look for variability if time-series photometric data exists.

As we have mentioned in the introduction, no time-series photometry for NGC 6981 has been published for the last $\sim$40 years, and none
of the time-series photometry that has been published was obtained using modern CCD imaging cameras. Our time-series observations provide more
data points ($\sim$100 compared to $\sim$20-60) over a longer time-base ($\sim$5~yr compared to $\sim$1-2~yr) than the photographic campaigns of previous authors,
which gives us the potential to detect variability on timescales from hours to months.
We also reach deeper by $\sim$4 magnitudes,
and achieve a much better photometric precision per data point ($\la$20~mmag down to $\sim$18.5~mag compared to $\sim$50~mmag at $\sim$17~mag),
which is mainly the consequence of the use of a larger telescope coupled with a sensitive CCD imager. Also, the technique of
difference imaging has enabled us to detect variables in the most crowded central parts of the globular cluster. Therefore, we are in a position to
fully revise the list of variables in the cluster.

In Table~\ref{tab:variables}, we present the details of all the confirmed variables in NGC~6981; namely, those variables for which we have detected brightness variations
in our light curves above the noise level, or which have clearly variable light curves published in the literature. The latter of these criteria covers the cases
of the two RR Lyrae variables V27 and V35 that lie outside of our field of view, and which therefore do not have a light curve in our data (see Section~\ref{sec:not_obs_var}).
We also provide a comprehensive set of celestial coordinates for all confirmed variables in our field of view in Table~\ref{tab:astrom},
and a set of finding charts in Figure~\ref{fig:finding_charts}, since clear finding charts displaying all confirmed variables
have not previously been published. This publication should therefore serve as the definitive reference for
the known variable star population of NGC~6981, and the reader should not need to refer elsewhere in order to collate the pertinent information
on these variables.

Our $V$, $r$ and $I$ time-series photometry for all the confirmed variables in NGC~6981 in our field of view is available in Table~\ref{tab:vri_phot}, of which we reproduce only a small
part in this paper, and which is available in full in electronic form (see Supporting Information).

The following sections describe our findings for the variables in NGC~6981, including our detection of previously unknown variables.

\subsection{Stars Without Light Curves}
\label{sec:not_obs_var}

The stars V27, V35, V38 and V42 do not have light curves in our data set. The reason for this is that V27 and V35 are not in our field of view,
and V42 is saturated in our reference image for each filter. For V38 we were unable to perform photometric measurements on the difference images
because the star lies very close to a saturated star in our reference image for each filter (see Section~\ref{sec:caveat}).
We note that both V27 and V35 have light curves published in \citet{dic1972b}, and V35 has a light curve published in \citet{nob1957}. These published light curves show
clear variations of the RR0 type, and periods for these variables
have been measured in more than one publication. Consequently we include these stars as confirmed variables in Table~\ref{tab:variables}.

The stars V38 and V42 have not been included in Table~\ref{tab:variables} because there have been no published light curves or periods for these
two stars, and the only claim to their variability are the brief statements of this property in \citet{saw1953} and \citet{dic1973}. The proximity
of V38 to a much brighter star may have confused its identification as a variable by \citet{saw1953}. Therefore, further time-series observations
will be required to promote either of these stars to confirmed variables.

\subsection{Stars That Do Not Show Variability}
\label{sec:fake_var}

Considering first the stars V1-V34 listed in SH1920, we find that our light curves for V6, V19, V22, V26, V30, V33 and V34 do not show any
variability, and that the $\sim$7-20~mmag RMS scatter in the light curves is consistent with the noise for the majority of constant stars of $\sim$17th magnitude
(see Figure~\ref{fig:rms_diagrams}). This confirms the comment made in SH1920 that V6, V19, V22, V26 and V33 ``do not appear to be conspicuously variable'', and
refutes the statement that V30 and V34 ``undoubtedly vary''. In the light of our precision photometry we have dropped these stars from the variables listed in
Table~\ref{tab:variables}.

Similarly, we find that V37, V39, V40 and V41, from the list of stars V35-V41 claimed to be variable by \citet{saw1953}, show no variability in our
light curves, although the light curves of V37 and V39 suffer from some outlier photometric measurements leading to artificially large RMS magnitude deviations
(see Figure~\ref{fig:rms_diagrams}). Hence, we have also dropped these stars from the variables listed in Table~\ref{tab:variables}.

Finally, for the variable star candidates S1-S9 and R1-R7 from \citet{kad1995}, we do not detect any variability above the $\sim$8-35~mmag light curve noise level
for the stars S1, S2, S4, S6, R1, R2, R5, R6 and R7, and therefore we have dropped these stars from the variables listed in Table~\ref{tab:variables}.
The remaining variable star candidates are RR Lyrae stars (see Section~\ref{sec:RRL_stars}).

\subsection{Detection Of New Variable Stars}
\label{sec:detect_var}

\begin{figure*}
\centering
\epsfig{file=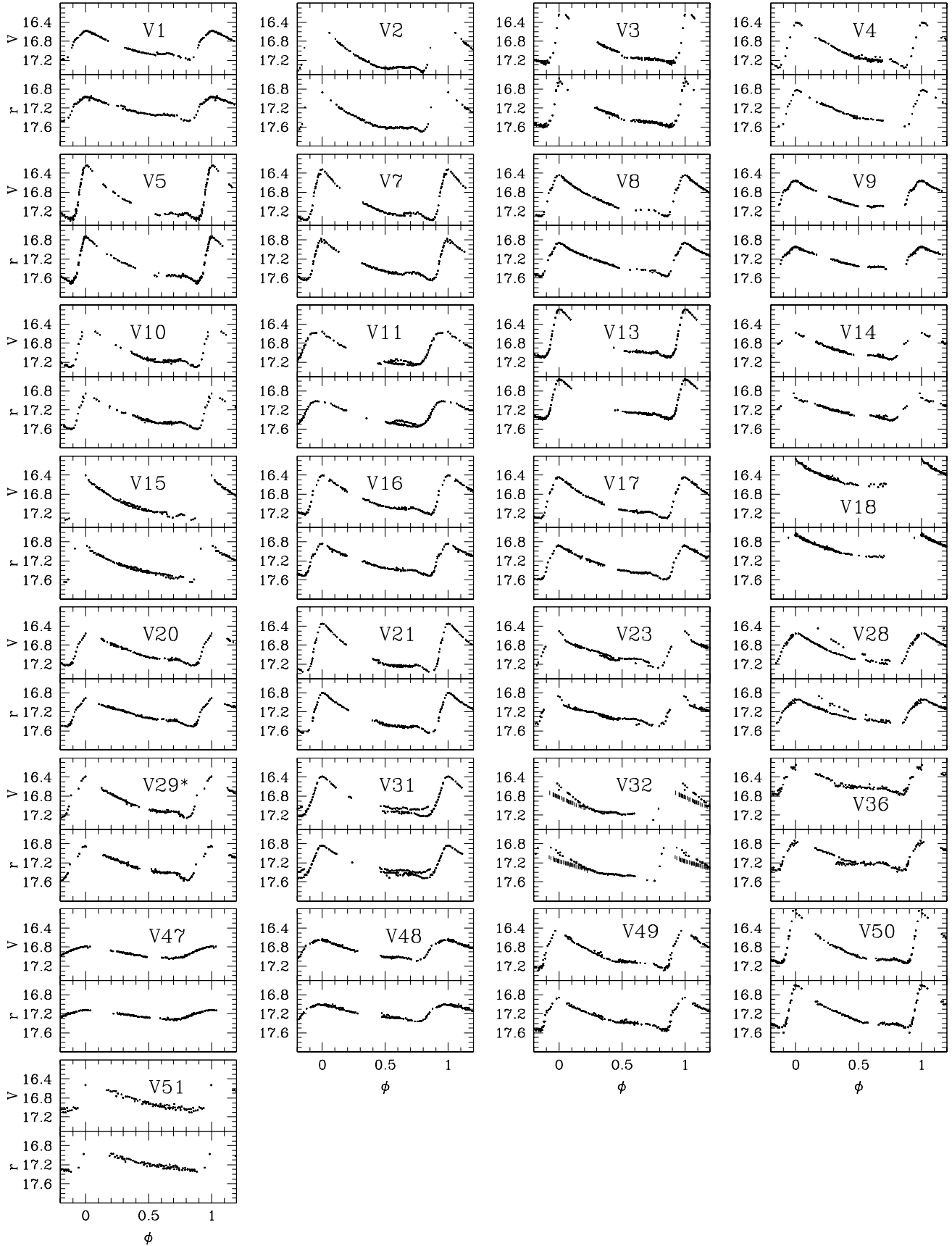,angle=0.0,width=0.98\linewidth}
\caption{Phased $V$ and $r$ light curves of the 29 RR0 stars for which we have data, using the periods listed in Table~\ref{tab:variables}.
         The light curve for V29, marked with an asterisk, is phased with the period and secular period change detailed in Table~\ref{tab:variables}.
         For V32, the only data to show the Blazhko effect are the observations from the nights in 2005, which do not phase well with the rest of the
         light curve, and are marked in the corresponding plot panel by vertical-dashed data points.
         \label{fig:RR0_phased}}
\end{figure*}

\begin{figure*}
\centering
\epsfig{file=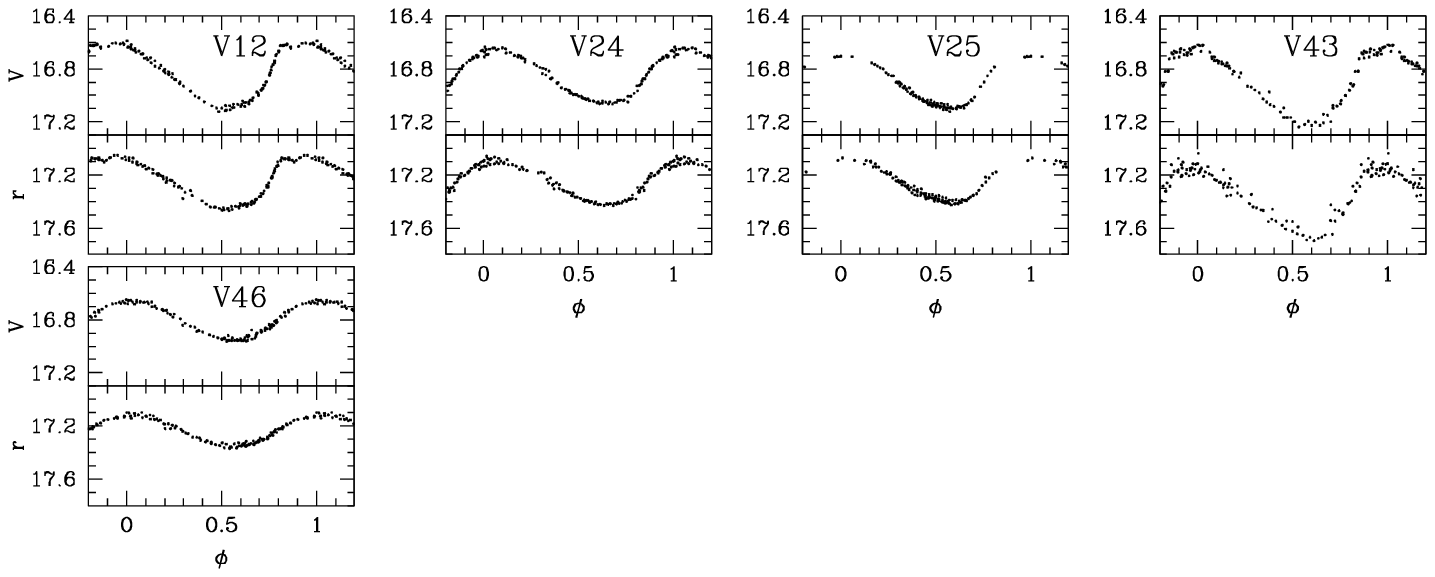,angle=0.0,width=0.98\linewidth}
\caption{Phased $V$ and $r$ light curves of the 5 RR1 stars for which we have data, using the periods listed in Table~\ref{tab:variables}.
         \label{fig:RR1_phased}}
\end{figure*}

Stars V43-V56 are newly detected variables. Of these, V43-V49 were discovered by inspection of the light curves of the variable star candidates from \citet{kad1995},
and V50-V53 were identified in the sequence of difference images as clearly varying sources. By converting each difference image $D_{kij}$ to an image of absolute deviations in units of
sigma $D^{\prime}_{kij}=|D_{kij}|/\sigma_{kij}$ (see Equation~\ref{eqn:noise_model}) and then constructing the sum of all such images $S_{ij}=\sum_{k}D^{\prime}_{kij}$ for each filter,
one can identify candidate variable sources as PSF-like peaks in the image $S_{ij}$. Using this method we discovered the new variables V54-V56 and we recovered all of the other
variable stars listed in Table~\ref{tab:variables}.

We also searched for variability in the set of light curves by applying the string-length method (\citealt{bur1970}; \citealt{dwo1983}) to each light curve
to determine the best period and a corresponding normalised string-length statistic $S_{\mbox{\scriptsize Q}}$. Exploration of the light curves with the smallest values
of $S_{\mbox{\scriptsize Q}}$ by phasing them
with their best period recovered most of the RR Lyrae stars in our list of variables, but did not yield any further variable star detections. In conjunction with the string-length
search method, we inspected the light curves of all stars with an unusually large RMS magnitude deviation using the plots in Figure~\ref{fig:rms_diagrams} as a guide to the typical
RMS values at a given magnitude. Again, we did not discover any more variable stars.

A useful result of the string-length period search was that we had obtained initial period estimates for the variable stars, from which we were able to determine accurate periods
by performing another string-length period search around the initial period estimates, but using a finer grid of test periods. Care was taken with the analysis of light curves
showing the Blazhko effect (see Section~\ref{sec:RRL_stars}) and in these cases we used appropriate subsets of self-consistent data to obtain the correct period.
Note that we were unable to determine periods for the variables V44, V45, V52 and V53, and that the period for V51 is unreliable (see Section~\ref{sec:RRL_stars}). For the
SX~Phoenicis stars V54-V56, we used a different method to determine their periods (see Section~\ref{sec:sxphe}).

\subsection{Variable Stars With Blended PSFs}
\label{sec:blended_var}

NGC~6981 has a very crowded central concentration of stars which implies that many of the variables we have observed are blended with other stars, and those that
we observed towards the centre of the globular cluster are almost certainly blended with one or more sources. In most cases the blend stars are much fainter than the variable
stars (which are dominated by bright RR Lyrae stars) and will have little effect on the measured reference fluxes and light curve variation amplitudes (see Section~\ref{sec:caveat}).
However, as a useful guide to the likelihood of a variable including a blend, in column 4 of Table~\ref{tab:variables} we indicate whether a variable star lies in the
crowded centre of the cluster or not, by setting a dividing line at a radius of 50~arcsec within which no sky background is visible in our reference images.

The PSF of all confirmed variable stars in the reference images was inspected in detail in order to identify blended PSFs. We find that V18, V24, V36,
V45, V51 and V52 are closely blended with stars of similar or greater brightness, and in these cases the variable star reference flux may be unreliable, depending on the success
of the deblending iterations applied by the {\tt DanDIA} software. The blend status of these variables is clearly marked in Table~\ref{tab:variables}.

\subsection{RR Lyrae Stars}
\label{sec:RRL_stars}

In Table~\ref{tab:variables} we present data on 36 confirmed RR Lyrae stars, 31 of which we classify as RR0 type, and 5 as RR1 type, although the classification for
V51 is a little uncertain. Classification was based primarily on light curve shape and period, and confirmation of our results was achieved by constructing the
Bailey diagram of amplitude versus period, which clearly separates the RR0 stars from the RR1 stars (see Figure~\ref{fig:bailey}). In the Bailey diagram, we also plot the average
distribution of the RR0 and RR1 variables in M3 (\citealt{cac2005}), the prototype Oosterhoff type I cluster, for comparison purposes. We note that V27 has an anomalously
large amplitude (taken from \citealt{dic1972b}) for its period, and that V46 has an anomalously small amplitude for its period.

Phased $V$ and $r$ light curves of the 29 RR0 and 5 RR1 stars for which we have
data are displayed in Figures~\ref{fig:RR0_phased}~\&~\ref{fig:RR1_phased}, respectively.
Of these variables, 5 RR0 and 2 RR1 are new discoveries (V47-V51 and V43,V46).

Our observations for V51 have poor phase coverage such that they only cover the falling part of the light curve. Consequently, we have been unable to derive
a reliable period, and the period $P=0.357335$~d that we present in Table~\ref{tab:variables} derived using the string-length method is probably a smaller period alias
of the real period. There is no information in the light curve that enables us to rule out other plausible period aliases. Although V51 is blended in the reference image,
it still lies neatly in the RR0 dominated region of the instability strip in the horizontal branch of the $V-r$ colour-magnitude diagram (CMD; see Figure~\ref{fig:CMD_Vr}).
Based on this evidence, V51 is clearly an RR Lyrae star, most likely of the RR0 type. Further evidence for the RR0 type comes from the non-sinusoidal shape of the light curve at the derived period.

It is currently believed that the Blazhko effect (\citealt{bla1907}), a periodic modulation of both the light curve amplitude and phase on timescales of typically tens to hundreds of days, occurs
frequently ($\ga$40~per~cent) in RR0 stars (\citealt{jur2009}; \citealt{kol2010}). Until recently the incidence rate of the Blazhko effect in RR0 stars was estimated to be substantially lower
(e.g. $\sim$25~per~cent in \citealt{smi1981}). However, this picture has become clearer with the execution and analysis of long-time-base quasi-continuous photometric observations of high precision
enabling one to detect smaller amplitude manifestations of the effect, and to detect the cases of RR0 stars exhibiting unstable Blazhko effects such as abrupt changes in the modulation amplitude
and/or the appearance/disappearance of the Blazkho effect. Note that the Blazhko effect has also been detected in a number of RR1 stars (e.g. \citealt{wil2008}).

\begin{figure}
\centering
\epsfig{file=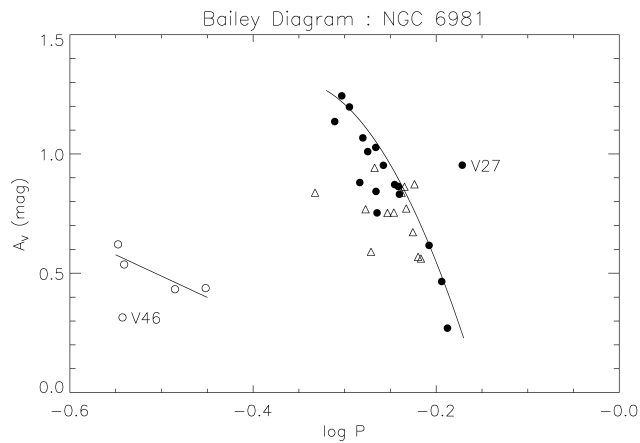,angle=0.0,width=\linewidth}
\caption{Bailey diagram of the light curve $V$ amplitude versus (log-)period for the RR Lyrae stars in NGC~6981. Variables of the RR0 type
         are plotted as solid circles where the light curve amplitude is well-defined between observed maxima and minima, and they
         are plotted as open triangles otherwise (which then represent a lower bound to the $V$ amplitude). Variables of the RR1 type
         are plotted as open circles. V51 has been omitted from this plot because it does not have a reliable period estimate.
         The continuous lines represent the average distribution of the RR0 and RR1 variables in M3 from \citet{cac2005}. The clear
         outliers in this diagram are labelled with their variable star designations, namely V27 and V46.
         \label{fig:bailey}}
\end{figure}

\begin{figure}
\centering
\epsfig{file=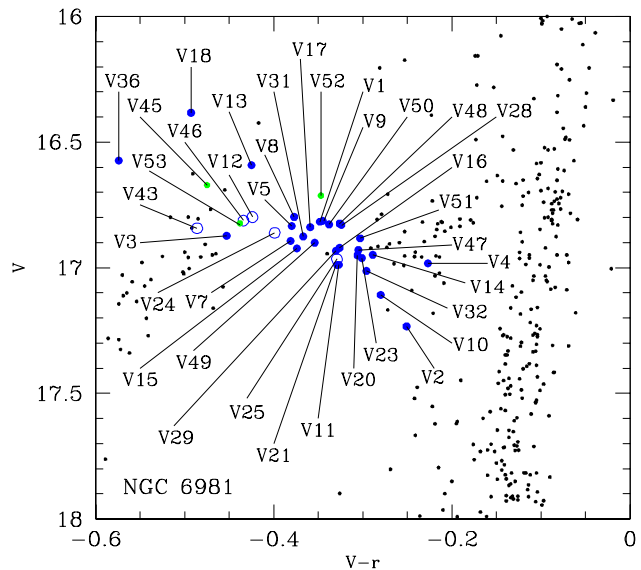,angle=0.0,width=\linewidth}
\caption{A portion of the $V-r$ CMD for NGC~6981, focussing on the horizontal branch and part of the red giant branch. RR Lyrae variables of the RR0 and RR1 type
         are plotted as solid and open blue circles, respectively. Three of the suspected RR Lyrae variables are plotted as solid green circles.
         \label{fig:CMD_Vr}}
\end{figure}

Our observations cover a $\sim$5~yr baseline with intense observations during small groups of nights. The $\sim$10~mmag precision that we have achieved at $V=\,$17~mag
allows us to detect small changes in amplitude and/or period of the RR Lyrae stars on the typical timescale of the Blazhko effect. The RR0 variables V11, V23, V28, V31 and V32
show a Blazhko effect that is clearly visible in the phased light curves in Figure~\ref{fig:RR0_phased}. Further careful inspection of the phased light curves (at a smaller scale and using
different plot symbol colours for each observation night) has allowed us to detect a small amplitude Blazhko effect in the light curves of the RR0 variables V10, V14, V15, V36 and V49, which
is not so visible in the phased light curves in Figure~\ref{fig:RR0_phased} due to the large scale that has been used. A more detailed analysis of the periodicity of the Blazhko effect
in these variables is not possible with our data since we have observed very few light curve maxima (see column 5 of Table~\ref{tab:variables}).
Note that we do not find evidence for the Blazhko effect in the light curves of the RR1 variables.

Having detected the Blazhko effect in 10 out of 29 RR0 variables, we can safely place a $\sim$34~per~cent lower limit on the rate of incidence of the Blazhko effect in RR0 stars in NGC~6981. We
state that this is a lower limit because we are convinced that with more observations during our time-base and/or with an extension to our time-base, 
it is possible to detect more small amplitude Blazhko effects.

We also investigated secular period changes for each RR Lyrae star for which the light curve does not phase as well as expected at the best-fit period given our photometric precision, which
included investigating the Blazhko effect variables discussed above. We used a modified version of the string-length method to search for the best fit of 
the following linear model for secular period change:
\begin{eqnarray}
&& \phi(t) = \frac{t - E}{P(t)} - \left \lfloor \frac{t - E}{P(t)} \right \rfloor  \\
&& P(t) = P_{0} + \beta ( t - E )                            \label{eqn:per_change}
\end{eqnarray}
where $\phi(t)$ is the phase at time $t$, $P(t)$ is the period at time $t$, $P_{0}$ is the period at the epoch $E$, and $\beta$ is the rate of period change.
For each light curve, the parameter space was searched at fixed epoch $E$, whose value is arbitrary, for the best-fit values of $P_{0}$ and $\beta$, using a brute force search for the minimum
string-length statistic of the light curve, and using a small range of periods around the previously determined best-fit period.

For the Blazhko effect variables described above,
it is clear that the secular period change model cannot explain the observed light curve variations. However, for the variable V29, which we originally suspected of
exhibiting the Blazhko effect, we find that the light curve produces a much better phased light curve for $P_{0}=\,$0.597472~d at epoch $E=\,$2453507.446~d
and with $\beta\approx-$1.38$\times$10$^{-8}$~d~d$^{-1}$, indicating that the star pulsation frequency is slowly increasing over time. The derived value for $\beta$ is rather
large compared to typical values for RR0 stars (e.g. \citealt{jur2001}) although it is certainly plausible.

In Figure~\ref{fig:CMD_Vr} we show a portion of the $V-r$ CMD for NGC~6981 that focusses on the horizontal branch and includes part of the red giant branch. As expected,
the RR0 stars cluster towards the red end of the instability strip, and the RR1 stars cluster towards the blue end. There is, however, some scatter in the positions of the
RR Lyrae stars, which is due to certain systematic errors in the measured mean magnitudes. Of the RR Lyrae stars with blended PSFs, V18 and V36 have clearly
over-estimated mean magnitudes in $V$ and $r$, which leads to a systematic error in their position on the CMD. The two other such stars, V24 and V51, do not suffer from
this problem, presumably because the {\tt DanDIA} software was able to perform successful deblending. Other outliers in the CMD may be explained by systematic
errors associated with calculating the mean magnitude for an incomplete phased light curve (e.g. V2, V13, V25).

\subsection{Suspected RR Lyrae Stars}
\label{sec:unknown}

\begin{figure}
\centering
\epsfig{file=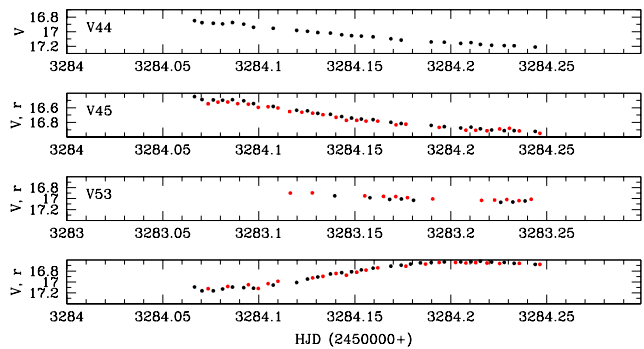,angle=0.0,width=\linewidth}
\caption{The $V$ (black circles) and $r$ (red circles) light curves for the variables V44 (top panel), V45 (upper-middle panel) and 
         V53 (lower two panels). The $r$ light curves have been offset in magnitude such that the mean $r$ magnitude matches the mean $V$ magnitude for each star.
         Period determination has not been possible due to the lack of data.
         \label{fig:lc_misc1}}
\end{figure}

Each of the variables V44, V45 and V53 lies close to a saturated star in the $V$ and $R$ reference images, and their light curves suffer from missing epochs
caused by the saturated pixels (see Section~\ref{sec:caveat}). In fact, for V44 and V45, the light curves only contain photometry from the night of 20041005
(the night of best-seeing), and V44 is further missing photometry in the $R$ filter. For V53, the light curve only contains photometry from the nights
of 20041004 and 20041005. It is this lack of data that has prohibited us from determining periods for these variables (see end of Section~\ref{sec:detect_var}).
The light curves for V44, V45 and V53 are presented in Figure~\ref{fig:lc_misc1}.

The variable V52 has a very poor phase coverage with our observations which has also prohibited us from determining a period. We present the light curve for this
variable in Figure~\ref{fig:lc_misc2}.

From the position of V45, V52 and V53 in the instability strip of the CMD in Figure~\ref{fig:CMD_Vr}, it is obvious that these variables are RR Lyrae stars. We
also believe that V45 is an RR0 type variable because the slow decline of the light curve over $\sim$0.2~d is not consistent with a sinusoidal light curve of the 
period expected for a typical RR1 type variable. We also believe that V44 is an RR0 type variable because it has a mean $V$ magnitude of $\sim$17.04~mag, typical of the RR Lyrae
stars in this cluster, and it exhibits a decline in brightness similar to V45 over the same time window (see Figure~\ref{fig:lc_misc1}).

\begin{figure}
\centering
\epsfig{file=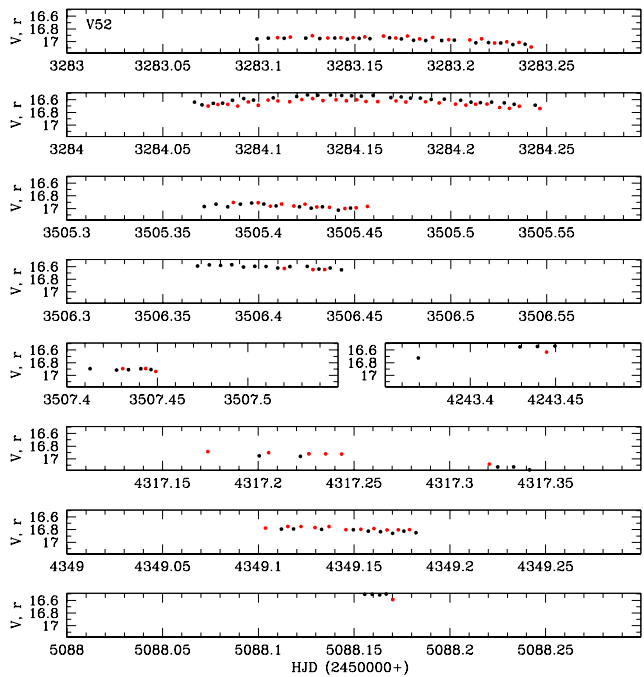,angle=0.0,width=\linewidth}
\caption{The $V$ (black circles) and $r$ (red circles) light curve for the variable V52. The $r$ light curve has been offset in magnitude such that the mean $r$ magnitude matches the mean $V$ magnitude.
         Period determination has not been possible due to the very poor phase coverage of our data.
         \label{fig:lc_misc2}}
\end{figure}

\subsection{SX Phoenicis Stars}
\label{sec:sxphe}

\begin{figure*}
\centering
\begin{tabular}{cc}   
\subfigure[]{\epsfig{file=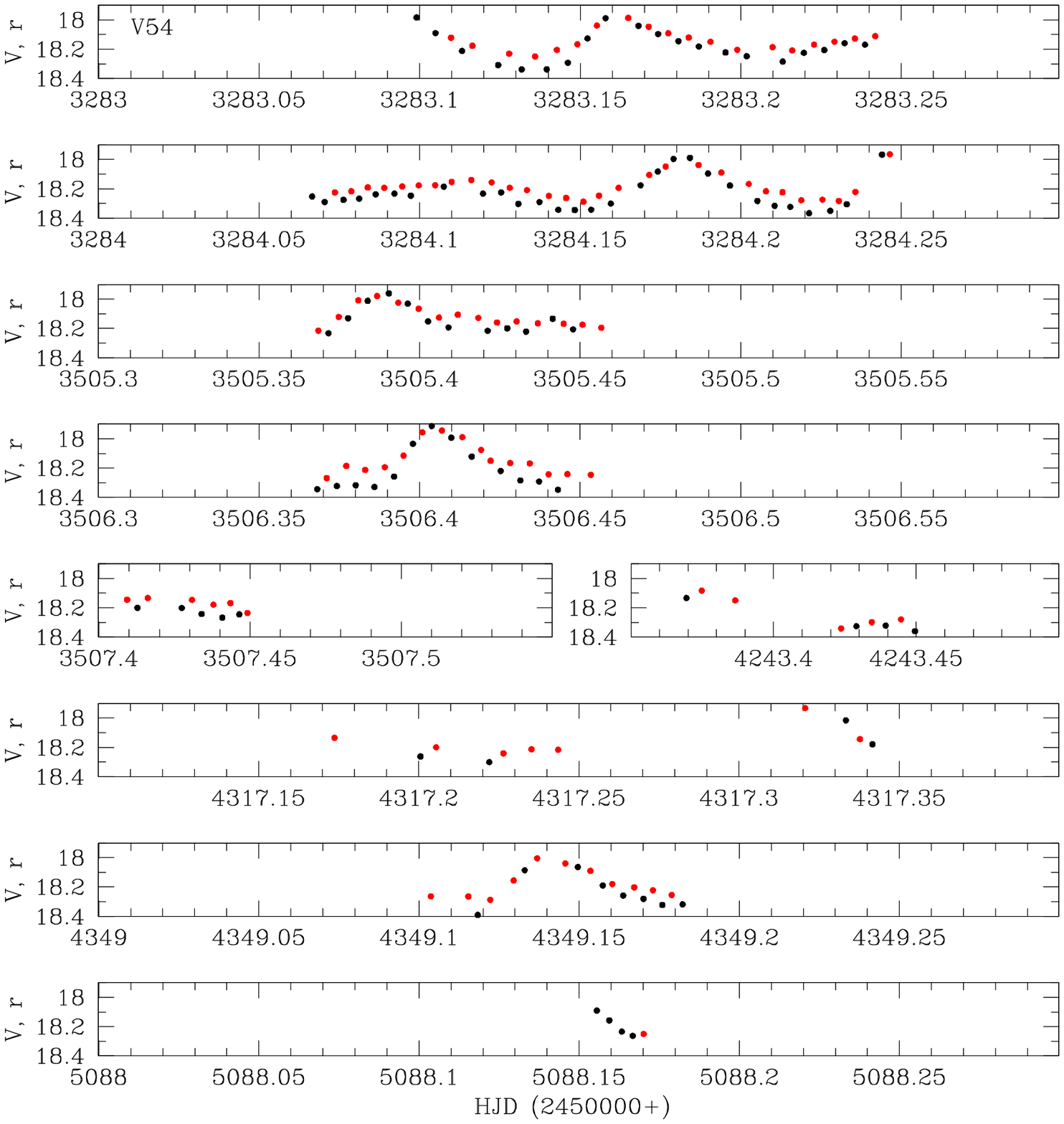,angle=0.0,width=0.48\linewidth} \label{fig:V54}} &
\subfigure[]{\epsfig{file=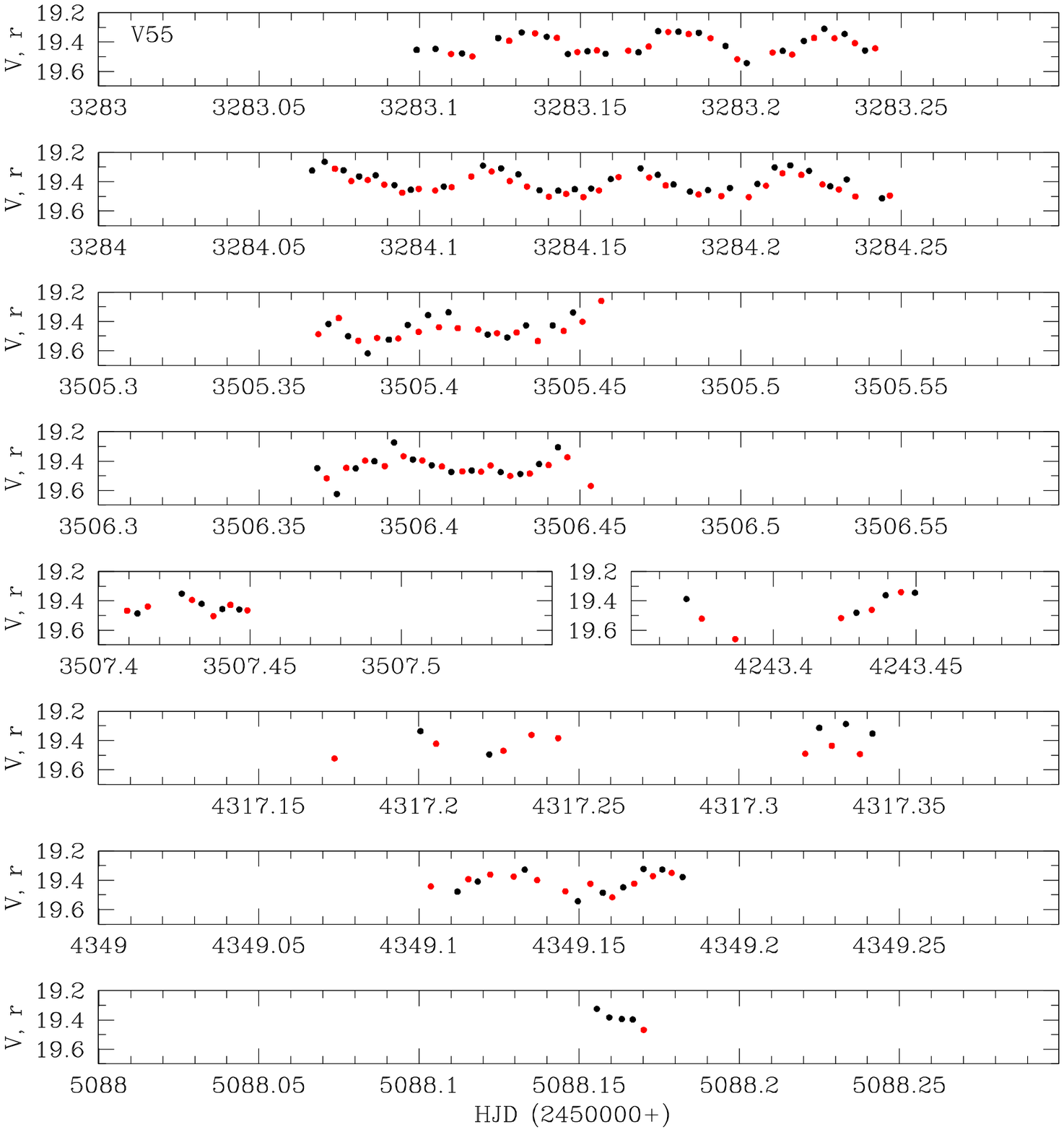,angle=0.0,width=0.48\linewidth} \label{fig:V55}} \\
\end{tabular}
\begin{tabular}{c}
\subfigure[]{\epsfig{file=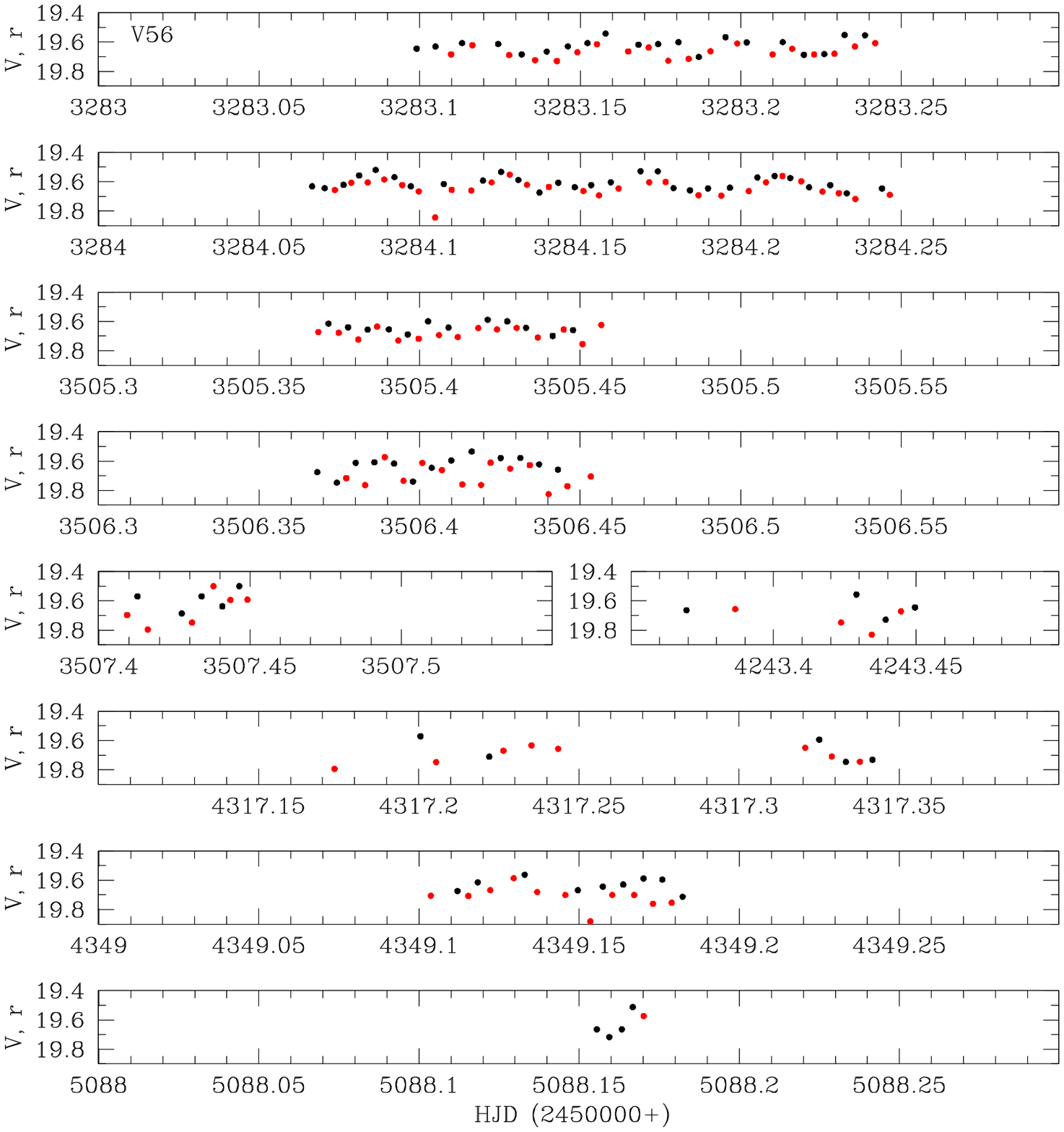,angle=0.0,width=0.48\linewidth} \label{fig:V56}} \\
\end{tabular}
\caption{The $V$ (black circles) and $r$ (red circles) light curves for the SX~Phe type variables: (a) V54, (b) V55, and (c) V56.
         The $r$ light curves have been offset in magnitude such that the mean $r$ magnitude matches the mean $V$ magnitude for each star.
         Mean photometric uncertainties per data point in $V$ and $r$ are 15 and 15~mmag for V54, 31 and 35~mmag for V55, and 40 and 51~mmag for V56,
         respectively.
         \label{fig:sxphe}}
\end{figure*}

Recently, a large number of SX~Phe type variables have been discovered in globular clusters in the blue straggler region of the CMD (e.g. \citealt{rod2000}; \citealt{jeo2003}; \citealt{dek2009}).
In our search for new variables in NGC~6981, we have discovered three SX~Phe type variables (V54-V56) and their light curves are presented in Figure~\ref{fig:sxphe}.
We analysed each $V$ and $r$ light curve separately
with the program {\tt PERIOD04} (\citealt{len2005}) in order to derive the dominant pulsation period and to search for any other pulsational frequencies. We accepted any frequencies with
oscillation amplitudes in $V$ that exceed the mean photometric uncertainties per data point in the corresponding light curve (15~mmag for V54, 31~mmag for V55, and 40~mmag for V56).
The results of the frequency analysis for the $V$ filter light curves are reported in Table~\ref{tab:sxphe}. For V54, we detect two significant frequencies with frequency ratio $f_{1}/f_{2}\approx\,$0.75,
which therefore correspond to the fundamental (F) and first overtone (1H) radial oscillation modes. For V55 and V56, we did not detect any significant frequencies apart from the dominant frequency,
and therefore we are unable to identify the mode of oscillation.
Note that the periods listed in Table~\ref{tab:variables} for these variables are the periods corresponding to the frequency of the largest amplitude oscillation.

The positions of the variables V54-V56 in the calibrated $V-I$ CMD are shown in Figure~\ref{fig:VI_cmd} as solid red circles. The $I$ images were observed at one epoch only,
and hence we have calculated the $V$ and $I$ mean magnitudes for each star from the three $V$ images taken closest in time to the three $I$ images, and from the
three $I$ images themselves, respectively. The $V$ and $I$ mean magnitudes differ in epoch by $\sim$18~min, which is not important for the constant stars, but
yields an approximate instantaneous colour for the variable stars. The fact that the $V$ and $I$
mean magnitudes are instantaneous rather than phase-averaged explains the spread of the RR Lyrae stars (solid and open blue circles) along the horizontal branch in
Figure~\ref{fig:VI_cmd} and the fact that the SX Phe variable V54 lies close to, but not within, the dashed box delimiting the blue straggler region, which
has been taken from \citet{har1993} for the cluster NGC~6366 and adapted to the distance and reddening of NGC~6981 (see Section~\ref{sec:distance}). The other two
variables V55 and V56 lie within the blue straggler region. This evidence, along with the fact that the periods of the largest
amplitude oscillation are in the range 0.03-0.08~d, confirms our SX~Phe type classification for these variables.

More observations of these variables are required to confirm the pulsational frequencies that we have detected, and to further characterise these stars.

\begin{table}
\caption{Detected pulsation frequencies for the SX~Phe variables discovered in NGC~6981. For each SX~Phe variable, we list
         the mean $V$ magnitude $A_{0}$ (column 2), the detected frequencies (columns 3 \& 4), the corresponding (full) amplitude $A_{V}$ in the $V$ filter (column 5),
         and the identified mode of oscillation at each frequency (column 6). The numbers in parentheses indicate
         the uncertainty on the last decimal place.
         }
\centering
\begin{tabular}{@{}ccccccc@{}}
\hline
Variable     & $A_{0}$   & Label   & Frequency      & $A_{V}$ & Mode \\
Star ID      & ($V$ mag) &         & (cyc~d$^{-1}$) & (mag)   &      \\
\hline
V54          & 18.191(1) & $f_{1}$ & 13.8915        & 0.246   & F    \\
             &           & $f_{2}$ & 18.4544        & 0.136   & 1H   \\
\hline
V55          & 19.392(3) & $f_{1}$ & 21.2618        & 0.174   & ---  \\
\hline
V56          & 19.613(3) & $f_{1}$ & 24.7099        & 0.087   & ---  \\
\hline
\end{tabular}
\label{tab:sxphe}
\end{table}

\section{Physical Parameters Of The RR Lyrae Stars}
\label{sec:RRL_physical}

\subsection{Fourier Light Curve Decomposition}
\label{sec:fourier}

\begin{figure}
\centering
\epsfig{file=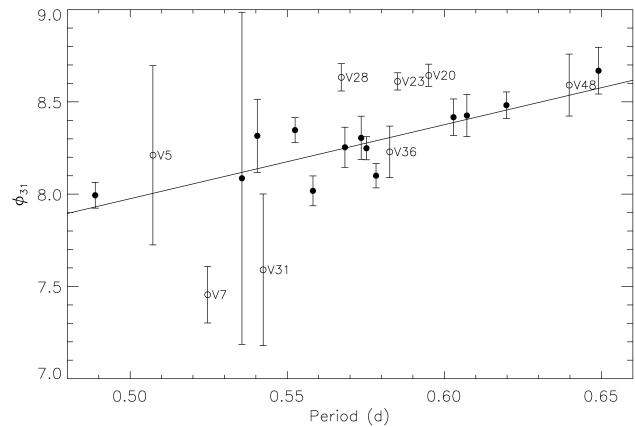,angle=0.0,width=\linewidth}
\caption{Plot of $\phi_{31}$ versus $P$ for the RR0 stars listed in Table~\ref{tab:fourier_coeffs}. Open circles with labels represent the ``outlier'' RR0 stars that
         were excluded from the fit of the cluster metallicity in Equations~\ref{eqn:RR0_metallicity} to Equations~\ref{eqn:conv_j_zw} (continuous line).
         \label{fig:phi31_v_p}}
\end{figure}

The Fourier light curve decomposition technique is based on fitting the following function to a light curve:
\begin{equation}
m(t) = A_{0} + \sum_{k=1}^{N} A_{k} \cos \left( \frac{2 \pi k}{P}(t - E) + \phi_{k} \right)
\label{eqn:fourier_coeff}
\end{equation}
where $m(t)$ is the magnitude of the object at time $t$, $P$ is the period, $E$ is the epoch, $A_{k}$ and $\phi_{k}$
are the $k$th Fourier coefficients, and $N$ is the degree (or number of harmonics) of the Fourier decomposition model.
The Fourier parameters (relative amplitudes and phase differences), which are independent of $E$, are subsequently defined by the relations:
\begin{eqnarray}
&& R_{ij} = A_{i} / A_{j}              \\
&& \phi_{ij} = j \phi_{i} - i \phi_{j}
\end{eqnarray}
for $1 \le i,j \le N$.

\citet{sim1981} demonstrated that the light curve Fourier parameters could provide useful information about pulsating variables, and
currently, for Cepheids and RR Lyrae stars, the parameters are routinely used to provide information about pulsation modes and resonances, and estimates of
physical parameters such as metallicity, luminosity and effective temperature. With this motivation, we have performed a Fourier decomposition of the $V$ filter light curves of the RR Lyrae stars
in NGC~6981 for which we have data and a reliable period estimate (i.e. 28 RR0 and 5 RR1 type variables). We note that the number of harmonics
required for a good fit of a given light curve depends as much on the quality of the data, such as the amount of scatter and phase coverage, as on the light curve shape.
Therefore we have chosen $N$ for each light curve based on the highest degree harmonic for which the amplitude is significant. However, for some variables, the corresponding
light curve does not yield a well constrained Fourier decomposition due to a lack of full phase coverage, and we have abandoned the fits for the RR0 type variables V2, V3,
V11, V13, V21, V29 and V32 due to this problem. For the RR0 variables exhibiting the Blazhko effect, we used the same subsets of
self-consistent data as those used for the period determination.

\begin{table*}
\caption{Fourier coefficients $A_{k}$ for $k=0,1,2,3,4$, and the Fourier parameters $\phi_{21}$, $\phi_{31}$ and $\phi_{41}$,
         for the 21 RR0 and 5 RR1 type variables for which the Fourier decomposition fit was successful. The numbers in parentheses indicate
         the uncertainty on the last decimal place. Also listed are the number of harmonics $N$ used to fit the light curve
         of each variable, and the deviation parameter $D_{\mbox{\scriptsize m}}$ (see Section~\ref{sec:RRL_metallicity}).
        }
\centering                   
\begin{tabular}{@{}lllllllllll@{}}
\hline
Variable      & $A_{0}$    & $A_{1}$   & $A_{2}$   & $A_{3}$   & $A_{4}$   & $\phi_{21}$ & $\phi_{31}$ & $\phi_{41}$ & $N$   & $D_{\mbox{\scriptsize m}}$ \\
Star ID       & ($V$ mag)  & ($V$ mag) & ($V$ mag) & ($V$ mag) & ($V$ mag) &             &             &             &       &                            \\
\hline
              &            &           &           &           & RR0 stars &             &             &             &       &                            \\
\hline
V1            & 16.926(3)  & 0.220(4)  & 0.101(3)  & 0.068(2)  & 0.033(3)  & 3.924(46)   & 8.482(72)   & 6.748(99)   & 6     & 2.7                        \\
V4            & 16.981(4)  & 0.311(6)  & 0.147(6)  & 0.109(5)  & 0.081(5)  & 3.963(44)   & 8.347(68)   & 6.357(88)   & 7     & 2.1                        \\
V5$^{a}$      & 16.969(32) & 0.355(41) & 0.167(46) & 0.150(35) & 0.098(23) & 4.480(359)  & 8.211(486)  & 7.022(667)  & 7     & 13.0                       \\
V7$^{a}$      & 17.036(13) & 0.334(21) & 0.204(17) & 0.152(13) & 0.063(13) & 3.762(110)  & 7.455(153)  & 5.602(215)  & 6     & 9.0                        \\
V8            & 16.966(6)  & 0.291(8)  & 0.141(11) & 0.101(10) & 0.063(6)  & 3.945(73)   & 8.254(109)  & 6.273(181)  & 8     & 2.2                        \\
V9            & 16.941(4)  & 0.236(5)  & 0.102(6)  & 0.072(5)  & 0.037(4)  & 3.958(67)   & 8.417(99)   & 6.670(147)  & 5     & 1.4                        \\
V10$^{b}$     & 16.971(10) & 0.302(17) & 0.143(9)  & 0.115(8)  & 0.075(6)  & 3.836(72)   & 8.018(81)   & 6.004(136)  & 8     & 1.3                        \\
V14$^{b}$     & 16.901(4)  & 0.223(7)  & 0.085(6)  & 0.048(5)  & 0.020(5)  & 3.959(54)   & 8.426(114)  & 7.065(232)  & 8     & 2.6                        \\
V15$^{b}$     & 16.987(11) & 0.314(15) & 0.169(15) & 0.105(14) & 0.074(11) & 4.105(134)  & 8.316(198)  & 6.234(276)  & 10    & 2.6                        \\
V16           & 16.905(3)  & 0.284(4)  & 0.134(5)  & 0.098(4)  & 0.060(4)  & 3.933(41)   & 8.249(63)   & 6.349(81)   & 7     & 0.5                        \\
V17           & 16.968(5)  & 0.295(3)  & 0.147(6)  & 0.103(7)  & 0.062(6)  & 4.004(84)   & 8.305(117)  & 6.392(170)  & 10    & 0.7                        \\
V18$^{c}$     & 16.376(58) & 0.296(97) & 0.115(60) & 0.056(58) & 0.031(39) & 3.979(731)  & 8.086(900)  & 6.278(1101) & 6     & 2.6                        \\
V20           & 16.934(2)  & 0.240(4)  & 0.113(4)  & 0.081(3)  & 0.046(2)  & 4.113(36)   & 8.644(60)   & 6.881(102)  & 6     & 1.4                        \\
V23$^{d}$     & 16.948(2)  & 0.241(3)  & 0.142(4)  & 0.109(4)  & 0.051(3)  & 4.177(34)   & 8.611(47)   & 6.678(82)   & 6     & 2.9                        \\
V28$^{d}$     & 16.968(4)  & 0.266(5)  & 0.133(5)  & 0.071(5)  & 0.031(4)  & 4.108(35)   & 8.633(74)   & 6.960(169)  & 8     & 1.9                        \\
V31$^{a,d}$   & 16.958(20) & 0.303(19) & 0.181(23) & 0.065(23) & 0.063(16) & 3.509(265)  & 7.590(411)  & 5.773(464)  & 5     & 8.5                        \\
V36$^{a,b,c}$ & 16.503(5)  & 0.214(6)  & 0.124(6)  & 0.062(5)  & 0.040(5)  & 3.746(76)   & 8.229(140)  & 6.266(179)  & 4     & 3.2                        \\
V47           & 16.926(1)  & 0.112(2)  & 0.030(1)  & 0.011(2)  & 0.001(2)  & 4.286(71)   & 8.669(126)  & 8.513(1108) & 4     & 2.0                        \\
V48$^{a}$     & 16.894(4)  & 0.176(5)  & 0.074(5)  & 0.038(3)  & 0.018(3)  & 3.955(92)   & 8.591(168)  & 6.935(241)  & 5     & 4.4                        \\
V49$^{b}$     & 16.937(3)  & 0.277(5)  & 0.144(4)  & 0.106(4)  & 0.068(4)  & 3.793(50)   & 8.100(66)   & 6.000(96)   & 5     & 2.7                        \\
V50           & 16.798(5)  & 0.358(8)  & 0.178(7)  & 0.129(5)  & 0.087(7)  & 3.882(47)   & 7.994(70)   & 5.842(86)   & 8     & 2.4                        \\
\hline
             &            &           &           &           & RR1 stars &             &             &             &       &                            \\
\hline
V12          & 16.856(2)  & 0.251(2)  & 0.047(3)  & 0.020(2)  & 0.018(2)  & 4.667(46)   & 2.594(122)  & 1.044(131)  & 7     & ---                        \\
V24$^{c}$    & 16.861(1)  & 0.206(2)  & 0.026(2)  & 0.013(2)  & 0.006(2)  & 4.598(80)   & 3.053(147)  & 2.714(342)  & 10    & ---                        \\
V25          & 16.882(3)  & 0.204(4)  & 0.030(5)  & 0.021(3)  & 0.009(3)  & 5.359(83)   & 3.726(205)  & 2.427(454)  & 5     & ---                        \\
V43          & 16.939(4)  & 0.278(5)  & 0.056(5)  & 0.014(5)  & 0.012(5)  & 4.628(98)   & 2.406(355)  & 0.702(414)  & 4     & ---                        \\
V46          & 16.806(1)  & 0.149(2)  & 0.009(2)  & 0.005(2)  & 0.003(2)  & 4.498(196)  & 3.519(367)  & 0.284(695)  & 4     & ---                        \\
\hline
\end{tabular}
\raggedright
\\ $^{a}$These RR0 variables have $D_{\mbox{\scriptsize m}} > 3$.
$^{b}$These RR0 variables exhibit a small amplitude Blazhko effect.
$^{c}$Closely blended with stars of similar or greater brightness.
$^{d}$These RR0 variables exhibit a large amplitude Blazhko effect.
\label{tab:fourier_coeffs}
\end{table*}

In Table~\ref{tab:fourier_coeffs}, we list the Fourier coefficients $A_{k}$ for $k=0,1,2,3,4$, and the Fourier parameters $\phi_{21}$, $\phi_{31}$ and $\phi_{41}$, for
the 21 RR0 and 5 RR1 type variables for which the Fourier decomposition fit was successful, along with the number of harmonics $N$ used to fit each light curve.

In previous papers (\citealt{are2008a}; \citealt{are2010}), we have discussed in detail the relations and corresponding zero points that may be
used to calculate the metallicities [Fe/H], absolute magnitudes $M_V$, effective temperatures $T_{\mbox{\scriptsize eff}}$, masses $M$ and radii $R$ of the RR~Lyrae
stars in globular clusters from the fit parameters derived from the Fourier decomposition of the $V$ light curves. Currently there is some debate about the reliability
of these relations with respect to the calculation of effective temperatures, masses and radii (e.g. \citealt{cac2005}; \citealt{cat2004}, see his Section~4).
Therefore, in the following sections, we refrain from calculating masses and radii, and for the other physical parameters we simply proceed to derive their values for
the RR Lyrae stars in NGC~6981 based on the published relations with little further discussion.

\subsection{Metallicity}
\label{sec:RRL_metallicity}

\citet{jur1996} derived the following empirical [Fe/H]$-\phi_{31}^{(s)}-P$ relation for RR0 type variables based on a data set of
$V$ light curves and independent spectroscopic metallicity estimates for 81 field RR0 variables:
\begin{equation}
\text{[Fe/H]}_{\mbox{\scriptsize J}} = -5.038 - 5.394 P + 1.345 \phi_{31}^{(s)}
\label{eqn:RR0_metallicity}
\end{equation}
where $\phi_{31}^{(s)}$ is a phase difference from the Fourier light curve decomposition using a sine series, as opposed to a cosine series, and
$P$ is the period (d). The value
of $\phi_{ij}^{(s)}$ may be obtained from the equivalent phase difference $\phi_{ij}$ in a cosine series via:
\begin{equation}
\phi_{ij}^{(s)} = \phi_{ij} - (i - j) \frac{\pi}{2}
\label{eqn:phase_rel}
\end{equation}
The metallicity [Fe/H]$_{\mbox{\scriptsize J}}$ may be transformed to the ZW (\citealt{zin1984}) metallicity scale [Fe/H]$_{\mbox{\scriptsize ZW}}$ by using the inverse of the relation from \citet{jur1995}:
\begin{equation}
\text{[Fe/H]}_{\mbox{\scriptsize J}} = 1.431 \text{[Fe/H]}_{\mbox{\scriptsize ZW}} + 0.88
\label{eqn:conv_j_zw}
\end{equation}

\begin{table}
\caption{Physical parameters of the RR0 variables. The numbers in parentheses indicate the uncertainty on the last decimal place.
        }
\centering  
\begin{tabular}{@{}lllll@{}}
\hline
Variable       & [Fe/H]$_{\mbox{\scriptsize ZW}}$ & $M_{V}$   & $\log(L/L_{\odot})$ & $T_{\mbox{\scriptsize eff}}$ \\
Star ID        &                                  & (mag)     &                     & (K)                          \\
\hline
V1             & $-$1.45(10)                      & 0.601(5)  & 1.669(2)            & 6334(30)                     \\
V4$^{c}$       & $-$1.33(9)                       & 0.623(8)  & 1.660(3)            & 6493(29)                     \\
V5$^{a}$       & $-$1.28(46)                      & 0.675(55) & 1.639(22)           & 6483(204)                    \\
V7$^{a}$       & $-$2.06(16)                      & 0.673(27) & 1.640(11)           & 6344(64)                     \\
V8             & $-$1.47(12)                      & 0.616(12) & 1.663(5)            & 6433(47)                     \\
V9             & $-$1.45(11)                      & 0.608(7)  & 1.666(3)            & 6358(41)                     \\
V10             & $-$1.66(10)                      & 0.630(21) & 1.657(8)            & 6408(36)                     \\
V14            & $-$1.46(13)                      & 0.598(9)  & 1.670(4)            & 6308(50)                     \\
V15            & $-$1.31(20)                      & 0.634(21) & 1.656(8)            & 6519(84)                     \\
V16            & $-$1.50(9)                       & 0.612(6)  & 1.664(2)            & 6408(26)                     \\
V17            & $-$1.44(13)                      & 0.606(7)  & 1.667(3)            & 6430(49)                     \\
V18$^{b}$      & $-$1.51(85)                      & 0.622(5)  & 1.660(49)           & 6441(372)                    \\
V20$^{a}$      & $-$1.21(9)                       & 0.622(5)  & 1.661(2)            & 6425(26)                     \\
V23$^{a,b,c}$  & $-$1.20(8)                       & 0.657(5)  & 1.646(2)            & 6454(21)                     \\
V28$^{a,b}$    & $-$1.11(10)                      & 0.622(7)  & 1.660(3)            & 6476(34)                     \\
V31$^{a,b}$    & $-$2.00(39)                      & 0.611(29) & 1.665(12)           & 6324(165)                    \\
V36$^{a,b}$    & $-$1.55(15)                      & 0.653(8)  & 1.648(3)            & 6366(57)                     \\
V47            & $-$1.39(14)                      & 0.641(3)  & 1.653(1)            & 6119(117)                    \\
V48$^{a}$      & $-$1.43(17)                      & 0.601(6)  & 1.669(3)            & 6294(69)                     \\
V49            & $-$1.65(9)                       & 0.623(7)  & 1.660(3)            & 6388(28)                     \\
V50            & $-$1.42(9)                       & 0.684(10) & 1.635(4)            & 6569(30)                     \\
\hline
Weighted Mean: & $-$1.48(3)                       & 0.623(2)  & 1.660(1)            & 6418(10)                     \\
\hline
\end{tabular}
\raggedright
\\ $^{a}$These variables are not included in the calculation of the mean metallicity (see Section~\ref{sec:RRL_metallicity}).
$^{b}$These variables are not included in the calculation of the mean absolute magnitude, log-luminosity or effective temperature (see Sections~\ref{sec:RRL_absmag}~\&~\ref{sec:RRL_teff}).
$^{c}$V4 and V23 have spectroscopic metallicity measurements of $-$1.61 and $-$1.28, respectively, on the ZW scale (see Section~\ref{sec:RRL_metallicity}).
\label{tab:RR0_properties}
\end{table}

\begin{table}
\caption{Physical parameters of the RR1 variables. The numbers in parentheses indicate the uncertainty on the last decimal place.
        }
\centering
\begin{tabular}{@{}lllll@{}}
\hline
Variable       & [Fe/H]$_{\mbox{\scriptsize ZW}}$ & $M_{V}$    & $\log(L/L_{\odot})$ & $T_{\mbox{\scriptsize eff}}$ \\
Star ID        &                                  & (mag)      &                     & (K)                          \\
\hline
V12            &  $-$1.59(21)                     &  0.568(9)  & 1.651(4)            & 7373(35)                     \\
V24            &  $-$1.77(27)                     &  0.587(10) & 1.643(4)            & 7280(32)                     \\
V25            &  $-$1.70(42)                     &  0.515(14) & 1.672(6)            & 7262(229)                    \\
V43            &  $-$1.63(59)                     &  0.601(23) & 1.638(9)            & 7372(74)                     \\
V46$^{a}$      &  $-$1.04(66)                     &  0.643(12) & 1.621(5)            & 7466(54)                     \\
\hline
Weighted Mean: & $-$1.66(15)                      &  0.568(6)  & 1.651(2)            & 7327(22)                     \\
\hline
\end{tabular}
\raggedright
\\ $^{a}$This variable is anomalous in the Bailey diagram (see Figure~\ref{fig:bailey}) and it has an anomalous $\phi_{31}$ value.
Therefore, we have not included this variable in the calculation of the mean metallicity, absolute magnitude, log-luminosity or effective temperature (see
Sections~\ref{sec:RRL_metallicity},~\ref{sec:RRL_absmag}~\&~\ref{sec:RRL_teff}).
\label{tab:RR1_properties}
\end{table}

The empirical relation defined in Equation~\ref{eqn:RR0_metallicity} is only applicable to light curves of RR0 variables
that are ``similar'' to the light curves that were used to derive Equation~\ref{eqn:RR0_metallicity}. For this purpose,
\citet{jur1996} defined a deviation parameter $D_{\mbox{\scriptsize m}}$ describing the deviation of a light curve
from the calibrating light curves, based on the Fourier parameter interrelations, and a compatibility condition that $D_{\mbox{\scriptsize m}} < 3$.
We list the values of $D_{\mbox{\scriptsize m}}$ for our light curves in column 11 of Table~\ref{tab:fourier_coeffs}.

In column 2 of Table~\ref{tab:RR0_properties}, we report the metallicity estimates for each RR0 variable along with their uncertainties as obtained
via Equation~4 from \citet{jur1996}. The metallicities show a large spread of values ranging over $\sim$1~dex due to a number of outliers. In order to highlight
which RR0 stars produce outlier metallicity values, we have plotted the parameters $\phi_{31}$ versus $P$ in Figure~\ref{fig:phi31_v_p} for the RR0 stars
in Table~\ref{tab:fourier_coeffs}, assuming that the uncertainty on $P$ is negligible. We find that V7, V20, V23 and V28 have clearly anomalous $\phi_{31}$ values given the
associated uncertainty. However, this is somewhat to be expected since V23 and V28 exhibit large amplitude Blazhko effects, and V7 has a non-compatible light curve with the relation
in Equation~\ref{eqn:RR0_metallicity} ($D_{\mbox{\scriptsize m}} > 3$). For V20, we can find no reason why the light curve may yield an anomalous $\phi_{31}$ value.

We have decided to ignore the $\phi_{31}$ values for the RR0 stars with non-compatible light curves (V5, V7, V31, V36, V48), for those exhibiting large amplitude Blazhko effects
(V23, V28, V31), and for V20. These ``outlier'' variables are plotted in Figure~\ref{fig:phi31_v_p} as open circles. We note that it is not necessary
to ignore the $\phi_{31}$ values for the blended RR0 stars since the $\phi_{31}$ parameter is a light curve shape parameter
that is unaffected by systematic errors in the light curve amplitude due to blending (see Section~\ref{sec:caveat}). We used the remaining RR0 
stars to fit a metallicity [Fe/H]$_{\mbox{\scriptsize ZW}}$ under the assumptions that the relation in Equation~\ref{eqn:RR0_metallicity} is correct and that NGC~6981 has no intrinsic spread in metallicity.
We obtain [Fe/H]$_{\mbox{\scriptsize ZW}} \approx -$1.48$\pm$0.03 and plot the corresponding relation as a straight line in Figure~\ref{fig:phi31_v_p}. Clearly the
\citet{jur1996} empirical [Fe/H]$-\phi_{31}^{(s)}-P$ relation holds for the RR0 variables in NGC~6981. It is also satisfying to note that only the stars V7, V20, V23 and V28 have 
$\phi_{31}$ values with residuals greater than 3$\sigma$.

For the RR1 variables, we used the empirical [Fe/H]$-\phi_{31}-P$ relation determined by \citet{mor2007} in order to estimate metallicities:
\begin{align}
\text{[Fe/H]}_{\mbox{\scriptsize ZW}} = \, & 52.466 P^{2} - 30.075 P + 0.131 \phi_{31}^{2} \notag \\
                                           & + 0.982 \phi_{31} - 4.198 \phi_{31} P + 2.424 \notag \\
\label{eqn:RR1_metallicity}
\end{align}
This relation provides metallicities on the ZW scale with a standard deviation in the residuals of 0.145~dex for the 106 cluster stars in the calibrating sample.
The [Fe/H]$_{\mbox{\scriptsize ZW}}$ values for the RR1 variables are reported in column 2 of Table~\ref{tab:RR1_properties} along with their
uncertainties calculated from the propagation of the $\phi_{31}$ uncertainties, but ignoring any uncertainty in the period.

We note that for the RR1 variable V46, the value of the $\phi_{31}$ parameter is anomalously large given the short period, and this has resulted in the
derivation of an anomalous metallicity of $-$1.04 for this star. In the Bailey diagram in Figure~\ref{fig:bailey}, we have already noted that V46 also has an
anomalously small amplitude for its period, and therefore we have decided to exclude this variable from the calculations of the mean physical
parameters of the RR1 variables in Table~\ref{tab:RR1_properties}. Consequently, we calculate a mean metallicity of $-$1.66$\pm$0.15 for the RR1 stars.

Finally, we note that the metallicities of only two RR Lyrae stars in NGC~6981 have previously been measured in the literature. \citet{smi1982} measured the
spectroscopic metallicity index $\Delta S$ for the RR0 type variables V4 and V23 as 7.6 and 5.5, respectively, which yield metallicities 
[Fe/H]$_{\mbox{\scriptsize ZW}} \approx -$1.61 and $-$1.28, respectively,  when converted to the ZW scale via the following relation from \citet{sun1991}:
\begin{equation}
\text{[Fe/H]}_{\mbox{\scriptsize ZW}} = -0.158 \Delta S - 0.408
\end{equation}
We include these metallicity measurements as a footnote to Table~\ref{tab:RR0_properties} for the sake of completeness.

\subsection{Absolute Magnitude And Luminosity}
\label{sec:RRL_absmag}

The following empirical relation yielding the absolute magnitude $M_{V}$ of the RR0 variables was determined from $V$ light curve data for 383 such cluster stars by \citet{kov2001}:
\begin{equation}
M_{V} = -1.876 \log P - 1.158 A_{1} + 0.821 A_{3} + K_{0}
\label{eqn:mv_rel_rr0}
\end{equation}
where $K_{0}$ is a constant to be determined. We adopt $K_{0} =\,$0.41~mag, as in \citet{are2010},
in order to be consistent with a true distance modulus for the LMC of $\mu_{0} =\,$18.5~mag (\citealt{fre2001}). We list the derived values of $M_{V}$ for the RR0 variables in
column 3 of Table~\ref{tab:RR0_properties}.

\citet{kov1998} derived a similar empirical relation for $M_{V}$ for RR1 variables from $V$ light curve data for 93 cluster stars:
\begin{equation}
M_{V} = -0.961 P - 0.044 \phi^{(s)}_{21} - 4.447 A_{4} + K_{1}
\label{eqn:mv_rel_rr1}
\end{equation}
where $K_{1}$ is a constant to be determined.
A value of $K_{1} =\,$1.261~mag was calculated by \citet{kov1998} to place Equation~\ref{eqn:mv_rel_rr1} on the Baade-Wesselink luminosity scale. However, \citet{cac2005} have
suggested that this value should be decreased to $K_{1} =\,$1.061~mag to make Equation~\ref{eqn:mv_rel_rr1} consistent with a LMC true distance modulus of $\mu_{0} =\,$18.5~mag.
We have therefore adopted $K_{1} =\,$1.061~mag, and we list the derived values of $M_{V}$ for the RR1 variables in column 3 of Table~\ref{tab:RR1_properties}.

The empirical relations in Equations~\ref{eqn:mv_rel_rr0}~\&~\ref{eqn:mv_rel_rr1} depend on a subset of the amplitude Fourier coefficients.
Therefore, any variable which suffers from a systematic over-estimate of the reference flux due to blending will yield Fourier amplitudes that are systematically too small, which
in turn leads to a systematic error in the calculated value for $M_{V}$. Of the RR Lyrae variables with Fourier decomposed light curves, V18, V24 and V36 are blended with stars of similar or
greater brightness, although only V18 and V36 have noticeably brighter mean magnitudes $A_{0}$ than the rest of the RR Lyrae stars (see Table~\ref{tab:fourier_coeffs}). Consequently,
we ignore the $M_{V}$ values for V18 and V36 in calculating the mean absolute magnitude in Table~\ref{tab:RR0_properties}. We also ignore the $M_{V}$ values for V23, V28 and V31 because
the strong Blazhko effect in these variables is likely to render the light curve amplitude measurements as unreliable. Finally, we note that the compatibility condition described
in Section~\ref{sec:RRL_metallicity} only applies to the empirical relation in Equation~\ref{eqn:RR0_metallicity}. Hence we calculate mean absolute magnitudes of
$M_{V} \approx\,$0.623$\pm$0.002 and 0.568$\pm$0.006~mag for the RR0 and RR1 variables, respectively.

We converted the absolute magnitudes listed in Tables~\ref{tab:RR0_properties}~\&~\ref{tab:RR1_properties} to log-luminosities $\log(L/L_{\odot})$ using:
\begin{equation}
\log \left(L/L_{\odot}\right) = -0.4 \left( M_{V} + B_{\mbox{\scriptsize C}}(T_{\mbox{\scriptsize eff}}) - M_{\mbox{\scriptsize bol},\odot} \right)
\label{eqn:RRL_logL}
\end{equation}
where we adopt a bolometric solar absolute magnitude of $M_{\mbox{\scriptsize bol},\odot} =\,$4.75~mag, and a bolometric correction $B_{\mbox{\scriptsize C}}(T_{\mbox{\scriptsize eff}})$
as a function of effective temperature from the calibrations for metal-poor stars from \citet{mon1998}. Anticipating our results from Section~\ref{sec:RRL_teff},
we use the mean effective temperatures of the RR0 and RR1 variables presented in Tables~\ref{tab:RR0_properties}~\&~\ref{tab:RR1_properties} to determine bolometric corrections
of -0.023 and 0.055~mag, respectively, from \citet{mon1998}, where we have reduced the mean effective temperature of the RR1 stars by 250~K in order to bring the mean temperature
in line with temperatures derived from observed $V-I$ colours (\citealt{are2010}).

\subsection{Effective Temperature}
\label{sec:RRL_teff}

The effective temperature $T_{\mbox{\scriptsize eff}}$ of RR0 and RR1 variables may also be estimated via the Fourier parameters of the $V$ light curves.
For the RR0 variables, we used the relations from \citet{jur1998} based on a comparison of empirical data to the \citet{kur1993} model atmospheres:
\begin{align}
(V - K)_{0} = \, & 1.585 + 1.257 P - 0.273 A_{1} - 0.234 \phi^{(s)}_{31} \notag \\
                 & + 0.062 \phi^{(s)}_{41} \label{eqn:v_minus_k} \\
\log T_{\mbox{\scriptsize eff}} = \, & 3.9291 - 0.1112 (V - K)_{0} - 0.0032 \text{[Fe/H]}_{\mbox{\scriptsize J}} \label{eqn:RR0_teff} \\
\notag
\end{align}
where [Fe/H]$_{\mbox{\scriptsize J}}$ is the metallicity calculated via Equation~\ref{eqn:RR0_metallicity}. We list the derived values
of $T_{\mbox{\scriptsize eff}}$ for the RR0 variables in column 5 of Table~\ref{tab:RR0_properties}.

Equation~\ref{eqn:v_minus_k} depends on the amplitude Fourier coefficient $A_{1}$, and therefore, for the same reasons given in
Section~\ref{sec:RRL_absmag}, we ignore the variables V18, V23, V28, V31 and V36 in the calculation
of the mean effective temperature for the RR0 variables $T_{\mbox{\scriptsize eff}} \approx\,$6418$\pm$10~K.

For the RR1 stars, one may use the following relation derived from the Fourier decomposition of theoretical light curves generated from hydro-dynamical pulsation models (\citealt{sim1993})
to estimate the effective temperature:
\begin{equation}
\log T_{\mbox{\scriptsize eff}} = 3.7746 - 0.1452 \log P + 0.0056 \phi_{31}
\label{eqn:RR1_teff}
\end{equation}
Note that the temperatures calculated from this equation are on a different absolute scale than the temperatures calculated from Equation~\ref{eqn:RR0_teff}.
We list the derived values of $T_{\mbox{\scriptsize eff}}$ for the RR1 variables in column 5 of Table~\ref{tab:RR1_properties}, and calculate a mean
effective temperature for the RR1 variables of $T_{\mbox{\scriptsize eff}} \approx\,$7327$\pm$22~K.

However, we urge caution in the use of the effective temperatures derived from the Fourier decomposition fit parameters since it has been
shown by \citet{cac2005} that the temperatures calculated via Equations~\ref{eqn:RR0_teff}~\&~\ref{eqn:RR1_teff} do not match the
colour-temperature relations predicted by the temperature scales of \citet{sek2000} or by the evolutionary models
of \citet{cas1999}. In our previous works on NGC~5466 (\citealt{are2008a}) and NGC~5053 (\citealt{are2010}), we also found similar systematic errors in the
effective temperatures derived from the light curve Fourier parameters for the RR1 stars. However, for the RR0 stars, we found (\citealt{are2010}) that the
effective temperatures derived from the light curve Fourier parameters agree with the colour-temperatures predicted by the horizontal branch models of \citet{van2006}.
Despite these caveats, we still provide the effective temperatures for the RR Lyrae
stars in NGC~6981 in order to enable comparisons with similar work for other globular clusters.

\section{Cluster Physical Parameters}
\label{sec:cluster_properties}

\begin{table*}
\caption{Metallicity estimates for NGC~6981 on the ZW scale as found from an extensive literature search.
        }
\centering
\begin{tabular}{lll}
\hline
[Fe/H]$_{\mbox{\scriptsize ZW}}$ & Reference       & Method                                                 \\
\hline
$-$1.48$\pm$0.03                 & This work       & Fourier light curve decomposition of the RR Lyrae stars \\
$-$1.46                          & \citet{mar2003} & Lick indices                                           \\
$-$1.41$\pm$0.30                 & \citet{tho2003} & Lick indices                                           \\
$-$1.26$\pm$0.07                 & \citet{gei1997} & Washington photometry                                  \\
$-$1.50$\pm$0.05                 & \citet{rut1997} & Ca II triplet                                          \\
$-$1.4$\pm$0.1                   & \citet{bro1996} & CMD analysis                                           \\
$-$1.40                          & \citet{har1996} & Globular cluster catalogue$^{a}$                       \\
$-$1.72$\pm$0.19                 & \citet{bro1990} & Absorption-line indices                                \\
$-$1.5                           & \citet{rod1990} & Calcium abundance                                      \\
$-$1.54$\pm$0.09                 & \citet{zin1984} & $Q_{39}$ spectral index                                \\
$-$1.58$\pm0.12$                 & \citet{zin1980} & $Q_{39}$ spectral index                                \\
\hline
\end{tabular}
\raggedright
\\ $^{a}$The catalogue version used is the updated 2003 version available at http://www.physics.mcmaster.ca/Globular.html.
\label{tab:cluster_met}
\end{table*}  

\subsection{Oosterhoff Classification}
\label{sec:type}

We calculate the mean periods for the RR0 and RR1 stars in NGC~6981 as $\left \langle P_{\mbox{\scriptsize RR0}} \right \rangle =\,$0.563~d and
$\left \langle P_{\mbox{\scriptsize RR1}} \right \rangle~=\,$0.308~d, respectively.
We have used the periods derived in this work for all RR0 and RR1 variables except V51 due to its unreliable period, and we included the periods from \citet{dic1972b} for the RR0
variables V27 and V35 which do not lie in our field of view. We also calculate the ratio of the number of RR1 to RR Lyrae type variables as
$n_{\mbox{\scriptsize RR1}} / \left( n_{\mbox{\scriptsize RR0}} + n_{\mbox{\scriptsize RR1}} \right) \approx\,$0.14 using 36 RR Lyrae stars with robust classifications
(i.e. excluding V44, V45, V52 and V53).
These quantities are clearly consistent with the classification of NGC~6981 as an Oosterhoff type~I cluster (\citealt{smi1995}), which agrees
with previous classifications (\citealt{van1973}; \citealt{cas1987}) based on the \citet{dic1972b} survey of RR Lyraes,
but employs an updated and more complete catalogue of RR Lyrae stars for a more robust result.

\subsection{Cluster Metallicity}
\label{sec:cluster_metallicity}

According to \citet{san2006}, the mean log-period of the field RR0 type variables is related to metallicity on the ZW scale as follows:
\begin{equation}
\left \langle \log P_{\mbox{\scriptsize RR0}} \right \rangle = -0.416 - 0.098 \text{[Fe/H]}_{\mbox{\scriptsize ZW}}
\label{eqn:mean_p_metallicity}
\end{equation}
\citet{cle2001} also previously observed that such a correlation exists for the Oosterhoff type I clusters, but that it breaks down for the
Oosterhoff type II clusters. However, since NGC~6981 is of Oosterhoff type I, we may employ Equation~\ref{eqn:mean_p_metallicity} 
to derive [Fe/H]$_{\mbox{\scriptsize ZW}} \approx~-$1.68 based on $\left \langle \log P_{\mbox{\scriptsize RR0}} \right \rangle \approx -$0.251~d
calculated for the same RR0 stars as in Section~\ref{sec:type}, and assuming that the relation also applies to cluster RR Lyraes.

The shortest and longest RR0 periods depend on the location of the blue and red edges, respectively, of the RR0 instability strip,
which \citet{san1993} showed also depend on the metallicity. \citet{san1993} quotes the following relations between metallicity and the
shortest $P_{\mbox{\scriptsize min}}$ and longest $P_{\mbox{\scriptsize max}}$ periods for globular cluster RR0 stars (his Figure~10):
\begin{eqnarray}
&& \log P_{\mbox{\scriptsize min}} = -0.526 - 0.117 \text{[Fe/H]}_{\mbox{\scriptsize ZW}} \label{eqn:min_p_metallicity} \\
&& \log P_{\mbox{\scriptsize max}} = -0.248 - 0.070 \text{[Fe/H]}_{\mbox{\scriptsize ZW}} \label{eqn:max_p_metallicity}
\end{eqnarray}
In NGC~6981, V2 and V27 are the RR0 variables with the shortest ($P_{\mbox{\scriptsize min}} =\,$0.465254~d) and longest known
($P_{\mbox{\scriptsize max}} =\,$0.673774~d) periods, respectively, from which we derive metallicities of $\sim-$1.66 and $\sim-$1.09 via Equations~\ref{eqn:min_p_metallicity}~\&~\ref{eqn:max_p_metallicity},
respectively. The updated version of Equation~\ref{eqn:min_p_metallicity} in \citet{san2006} also yields a metallicity of $\sim-$1.66 after taking into account
the period shift of the short-period locus between field and cluster RR0 variables.

The large discrepancy between the two metallicity values we have derived from considering the location of the blue and red edges of the instability strip would be reduced
if any of the variables V44, V45, V51, V52 and V53 are found to be RR0 stars with periods shorter than 0.465~d or longer than 0.674~d. In fact, Equation~\ref{eqn:max_p_metallicity} is the most
sensitive to a change in the period boundaries, and all that is required to bring the metallicity estimates from the shortest and longest periods into agreement at
[Fe/H]$_{\mbox{\scriptsize ZW}} \sim -$1.66 is a $\sim$10~per~cent increase in the value of $P_{\mbox{\scriptsize max}}$ to $\sim$0.74~d. We also note that any future periods
determined for V44, V45, V51, V52 and V53 will influence the mean (log-)period of the RR0 variables in NGC~6981, which may also have a substantial effect on the
metallicity estimated via Equation~\ref{eqn:mean_p_metallicity} due to the high sensitivity of this relation on the mean log-period.
In fact, we remain rather sceptical of the metallicity estimates for NGC~6981 that we have derived via Equations~\ref{eqn:mean_p_metallicity}, \ref{eqn:min_p_metallicity}~\&~\ref{eqn:max_p_metallicity}
from the work of Sandage for just these reasons, and Figure~4 from \citet{cac2005} provides a clear graphical illustration of the potential problems with this method that we have described.

We have compiled estimates of [Fe/H] on the ZW scale given in the literature for NGC~6981 and obtained using a variety of methods, and we summarise the results in
Table~\ref{tab:cluster_met}, which includes references and a very brief description of the methods employed.

\begin{table*}
\caption{True distance moduli (column 1), and distance estimates (column 2), for NGC~6981 as found from an extensive literature search.
        }
\centering  
\begin{tabular}{llll}
\hline
$\mu_{0}$ (mag)   & Distance (kpc) & Reference         & Method                                            \\
\hline
16.117$\pm$0.047  & 16.73$\pm$0.36 & This paper        & Fourier light curve decomposition of the RR0 stars \\
16.111$\pm$0.047  & 16.68$\pm$0.36 & This paper        & Fourier light curve decomposition of the RR1 stars \\
16.42$\pm$0.01    & 19.23$\pm$0.09 & \citet{cas2001}   & Magnitude of the zero-age horizontal branch       \\
16.52             & 20.14          & \citet{zoc2000}   & Globular cluster luminosity function              \\
16.08             & 16.44          & \citet{fer1999}   & Magnitude of the zero-age horizontal branch       \\
---               & 17.0           & \citet{har1996}   & Globular cluster catalogue$^{a}$                  \\
16.30$\pm$0.05    & 18.20$\pm$0.42 & \citet{jim1996}   & CMD analysis                                      \\
---               & 16.4           & \citet{zin1980}   & $Q_{39}$ spectral index                           \\
16.29             & 18.1           & \citet{sea1978}   & Spectral indices                                  \\
16.30             & 18.2           & \citet{wol1975}   & Mean magnitude of the RR Lyrae stars              \\
15.73             & 14.0           & \citet{dic1972b}  & Mean magnitude of the RR Lyrae stars              \\
17.0              & 25             & \citet{kro1960}   & Mean magnitude of the RR Lyrae stars              \\
\hline
\end{tabular}
\raggedright  
\\ $^{a}$The catalogue version used is the updated 2003 version available at http://www.physics.mcmaster.ca/Globular.html.
\label{tab:cluster_dist}
\end{table*}

In Section~\ref{sec:RRL_metallicity}, we calculated mean metallicities of $-$1.48$\pm$0.03 and $-$1.66$\pm$0.15 on the ZW scale for the RR0 and RR1 type variables, respectively.
The smaller uncertainty on the result for the RR0 stars is mainly a consequence of the inclusion of more stars in the calculation of the mean metallicity for the RR0 variables (13 RR0
stars as opposed to 4 RR1 stars). It should be noted that the uncertainties we have quoted only represent the internal error for each metallicity estimate, and do not include
any systematic errors that may be inherent in the Fourier light curve decomposition method that we have employed. Given the fact that different empirical relations have been used 
to estimate the metallicities of the RR0 and RR1 stars, there may be some systematic offset between the metallicity estimates of the two types of variable.
However, assuming that this is not the case, and performing a combined mean metallicity estimate
for all 17 stars also yields a mean metallicity of $\sim -$1.48$\pm$0.03. Hence we adopt [Fe/H]$_{\mbox{\scriptsize ZW}} \approx -$1.48$\pm$0.03 as our estimate of the metallicity
of NGC~6981 on the ZW scale via the Fourier decomposition of the light curves of the RR Lyrae variables, which lies in the middle of the distribution of the other metallicity
estimates for the cluster from a large range of independent methods that are listed in Table~\ref{tab:cluster_met}.

Recently, \citet{car2009} collected a homogeneous set of intermediate to high-resolution spectra of $\sim$2000 red giant branch stars in 19 Galactic globular clusters and used them
to define an accurate and updated metallicity scale. They also provide the following transformation from the ZW scale to this new metallicity scale:
\begin{equation}
\text{[Fe/H]}_{\mbox{\scriptsize UVES}} = -0.413 + 0.130 \text{[Fe/H]}_{\mbox{\scriptsize ZW}} - 0.356 \text{[Fe/H]}_{\mbox{\scriptsize ZW}}^{2}
\end{equation}
where [Fe/H]$_{\mbox{\scriptsize UVES}}$ represents the metallicity on the new \citet{car2009} scale. Our estimate of the metallicity of NGC~6981 transformed onto this new scale
is [Fe/H]$_{\mbox{\scriptsize UVES}} \approx -$1.38$\pm$0.03. This value is consistent to within the uncertainties with the metallicity that \citet{car2009} derive for NGC~6981 of
[Fe/H]$_{\mbox{\scriptsize UVES}} \approx -$1.48$\pm$0.07, which is based on a recalibration of the $Q_{39}$ and $W^{\prime}$ spectral indices, and a weighted mean of the metallicity
estimates available on different scales, but transformed to the new scale.

\subsection{Distance}
\label{sec:distance}

To determine the distance to NGC~6981, the RR Lyrae stars can serve as standard candles since we have estimated their mean absolute magnitudes in Section~\ref{sec:RRL_absmag}
as $M_{V} \approx\,$0.623$\pm$0.002 and 0.568$\pm$0.006~mag for the RR0 and RR1 variables, respectively. We calculate the average of the RR0 and RR1 star mean $V$ magnitudes
($A_{0}$ from Table~\ref{tab:fourier_coeffs}) as $\sim$16.927$\pm$0.001 and $\sim$16.865$\pm$0.001~mag, respectively (excluding the variables V18, V23, V28, V31 and V36 from the
calculation for the RR0 stars because of the likely influence of blending and the Blazhko effect on the measured mean
magnitudes for these stars, and excluding the variable V46 from the calculation for the RR1 stars because of its anomalous status).
This implies mean apparent distance moduli of $\mu \approx\,$16.303$\pm$0.002 and 16.297$\pm$0.006~mag for the RR0 and RR1 variables, respectively.

NGC~6981 is known to be a globular cluster with relatively low interstellar extinction. The many values in the literature for the associated interstellar reddening range from
$E(B-V) =\,$0.00~mag (\citealt{sea1978}) to $E(B-V) =\,$0.11$\pm$0.03~mag (\citealt{rod1990}), with a median value of $E(B-V) =\,$0.06~mag. The two most 
recent determinations include $E(B-V) =\,$0.06~mag from the analysis of 100~$\mu$m emission from interstellar dust (\citealt{sch1998}), and
$E(B-V) =\,$0.070$\pm$0.015~mag from the comparison of predicted and measured colours as a function of $B$ light curve amplitude for the RR0 stars (\citealt{des1999}).
For the purpose of our analysis, we adopt a reddening of $E(B-V) =\,$0.06$\pm$0.015~mag. 

Assuming $R_{V} =\,$3.1 for our Galaxy, we derive mean true distance moduli of $\mu_{0} \approx\,$16.117$\pm$0.047 and 16.111$\pm$0.047~mag for the RR0 and RR1 variables, respectively,
which translate to mean distances of $\sim$16.73$\pm$0.36~kpc and $\sim$16.68$\pm$0.36~kpc, respectively. Even though both of these distance estimates are consistent
with a true distance modulus for the LMC of $\mu_{0} =\,$18.5~mag, they are calculated via two different empirical relations, each of which has its own systematic 
uncertainty, and
therefore we cannot simply calculate an overall distance estimate from all RR Lyrae stars put together.

In Table~\ref{tab:cluster_dist}, we report our distance estimates for NGC~6981 along with those we found in the literature search. We list both true distance moduli (where available)
and distances (kpc) along with references and methods employed. The distances from the literature show considerable spread, although varying assumptions about the amount of reddening
towards NGC~6981 certainly contribute significantly to the scatter. Our distance estimates would range from $\sim$15.5~kpc to $\sim$18.2~kpc by simply changing our assumption about
the reddening to values in the range 0.00$\,\le E(B-V) \le\,$0.11 (defined by the minimum and maximum measured reddening values in the literature). We observe that our
distance estimates for NGC~6981 fall towards the shorter end of the distribution of distance estimates from the literature, and it is satisfying to note that 
our mean distances for both types of RR Lyrae star are in excellent agreement.

\begin{table*}
\caption{Age estimates for NGC~6981 as found from an extensive literature search.
        }
\centering   
\begin{tabular}{lll}
\hline
Age (Gyr)                          & Reference         & Method                               \\
\hline
12.75$\pm$0.75                     & \citet{dot2010}   & Isochrone fitting to the CMD         \\
9.5                                & \citet{mei2006}   & Isochrone fitting to the CMD         \\
12.6$\pm$1.0                       & \citet{rak2005}   & Str\"omgren photometry                \\
14                                 & \citet{zoc2000}   & Globular cluster luminosity function \\
12.7$\pm$2, 13.0$\pm$2, 13.5$\pm$2 & \citet{jim1996}   & CMD analysis                         \\
12.4$\pm$2.2                       & \citet{sar1989}   & CMD analysis                         \\
\hline
\end{tabular}
\label{tab:cluster_ages}
\end{table*}

\subsection{Age}
\label{sec:age}

It is only in the last 20 years that measurements of the age of NGC~6981 have appeared in the literature, and they number far fewer then the number of measurements of the
metallicity and distance. We present the measured ages from the literature in Table~\ref{tab:cluster_ages} and we observe that the measurements have converged to a
value of $\sim$13~Gyr.

An exquisitely precise and deep CMD for NGC~6981 has been constructed by \citet{and2008} through careful processing of a stack of images of NGC~6981
obtained with the Wide-Field Channel of the Advanced Camera for Surveys (ACS) on board the {\it Hubble Space Telescope}. These images were observed as part of
the ACS Survey of Globular Clusters which includes observations of the central $\sim$3$\times$3~arcmin$^{2}$ of 65 Galactic globular clusters. Subsequent
analysis of all the cluster CMDs in the survey via main sequence fitting and main-sequence turnoff (MSTO) measurements allowed \citet{mar2009} to determine precise
(2\%-7\% uncertainty) relative ages between the clusters. \citet{dot2010} placed these relative ages on an absolute scale, thereby deriving an age of 12.75$\pm$0.75~Gyr
for NGC~6981, which is the most precise age measurement to date.

Our calibrated photometric data for NGC~6981 in the $V$ and $I$ bands does not reach as deep as the corresponding ACS data. Also, in our observations, most stars
in the central region (3$\times$3~arcmin$^{2}$) of the globular cluster are blended with other cluster members, making accurate absolute photometry a very difficult task for
the vast majority of cluster members, and introducing systematic photometric errors due to poorly modelled PSFs and blended light contamination. The space-based observations
enjoy a much better spatial resolution than our data due to the lack of blurring from an intervening atmosphere, and therefore the ACS photometry is of much better
accuracy with considerably less systematic errors due to blending. Hence, any analysis of the CMD from our data for determining the cluster age cannot compete with the similar analysis
by \citet{dot2010} of the ACS CMD.

In the calibrated $V-I$ CMD presented in Figure~\ref{fig:VI_cmd}, we overplot a theoretical isochrone (solid black line) corresponding to the \citet{dot2010} age of 12.75~Gyr, our derived
metallicity of [Fe/H]$_{\mbox{\scriptsize ZW}} = -$1.48, an $\alpha$-element abundance of [$\alpha$/Fe]$ = +0.3$ (a typical assumption for globular clusters), and an apparent distance modulus
of $\mu = 16.303$~mag (our result for the RR0 stars). The isochrone has been linearly interpolated from the four nearest isochrones in age and metallicity as taken from the stellar evolutionary
models of \citet{van2006}, which are also plotted in Figure~\ref{fig:VI_cmd} as solid coloured lines. The isochrones show that our CMD is consistent with the age for NGC~6981
derived by \citet{dot2010}, although the fit to the upper part of the red-giant branch is too red by about $\sim$0.1~mag. This mismatch can be slightly improved
by adopting a more metal-poor isochrone but at the expense of a worsening fit near the main sequence turn-off. Note that we have not applied a $V-I$ colour correction
to the isochrones due to reddening since this correction makes the isochrone fit considerably worse by shifting the isochrones by $\sim$0.076~mag to the right (assuming $E(B-V) =\,$0.06~mag).
This implies that the adopted reddening is too large and/or that the adopted theoretical isochrones are systematically too red at the cluster age and metallicity.

\begin{figure}
\centering
\epsfig{file=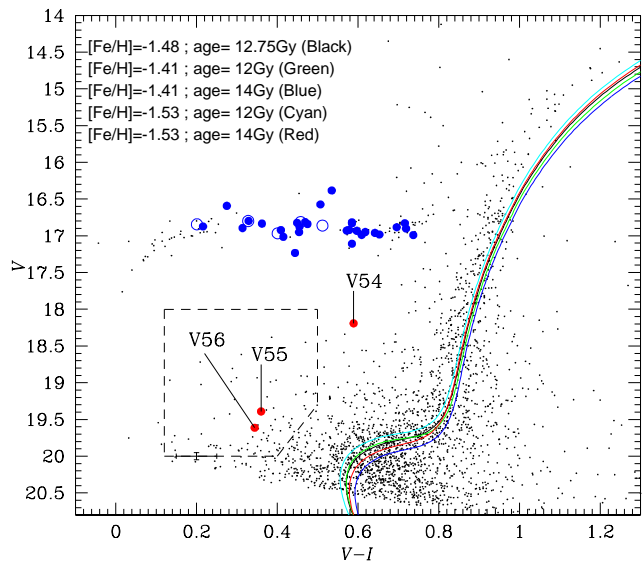,angle=0.0,width=\linewidth}
\caption{The $V-I$ CMD for NGC~6981. Note that the magnitude measurements approximate instantaneous magnitudes which
         explains the spread in the positions of the RR Lyrae stars along the horizontal branch (solid and open blue circles for the RR0 and RR1 variables, respectively)
         and the fact that one of the SX Phe variables (solid red circles) does not lie within the dashed box delimiting the blue straggler region. Theoretical isochrones
         from the stellar evolutionary models of \citet{van2006} are plotted as solid coloured lines. The black solid isochrone has been linearly interpolated from the
         other isochrones to the \citet{dot2010} age of 12.75~Gyr and our derived metallicity of [Fe/H]$_{\mbox{\scriptsize ZW}} = -$1.48.
         \label{fig:VI_cmd}}
\end{figure}

\section{Summary And Conclusions}
\label{sec:conclusions}

We employed the technique of difference image analysis to perform precision differential photometry on a set of CCD time-series observations of the globular cluster NGC~6981
in the Johnson $V$, $R$ and $I$ filters. The difference imaging technique has allowed us to study in detail the variable star population of the cluster, including in
the highly crowded central regions where traditional PSF fitting fails. Compared to previous photographic studies of this cluster, our photometry is deeper by $\sim$4~magnitudes,
achieves a much better precision per data point ($\le$20~mmag down to $\sim$18.5~mag as opposed to $\sim$50~mmag at $\sim$17~mag), and covers a longer time-base of $\sim$5~years
with substantially more data points ($\sim$100 compared to $\sim$20-60).

Consequently, we have been able to perform a census of the variable star population in NGC~6981, and to clarify the status of already known and suspected variables in the cluster.
We have shown that 20 of the 58 stars that are labelled as variables in the literature are actually non-variable, and we present the details of all 43 confirmed variables
in NGC~6981 in Table~\ref{tab:variables}.

We have discovered 14 of the 43 confirmed variables in NGC~6981 from our data. The new detections consist of 11 RR Lyrae variables and 3 SX Phe variables, although we
were only able to derive reliable periods for 6 of the newly discovered RR Lyrae stars. For the 27 of the 29 previously known RR Lyrae variables that lie in
our field of view, we calculated a new set of ephemerides that improve substantially on their previous values in terms of accuracy.

The current set of 43 variables in NGC~6981 consists of 40 RR Lyrae stars and 3 SX Phe stars. Furthermore, the set of RR Lyrae stars is made up of 31 RR0 type variables, 5 RR1 type variables,
and 4 RR Lyrae variables with an ambiguous subtype, although these variables are most likely to be of the RR0 type (V44, V45, V52 and V53). Based on the mean
periods for the RR0 and RR1 variables, and the ratio of the number of RR1 to RR0 type variables, we confirm the Oosterhoff type I classification for NGC~6981.
We report the detection of a strong Blazhko
effect in 5 RR0 variables (V11, V23, V28, V31 and V32), and a smaller amplitude Blazhko effect in another 5 RR0 variables (V10, V14, V15, V36 and V49), which implies a $\sim$34~per cent
lower limit on the rate of incidence of the Blazhko effect in RR0 stars in NGC~6981. We also find that the RR0 star V29 exhibits a secular period change of 
$\beta\approx-$1.38$\times$10$^{-8}$~d~d$^{-1}$, indicating that the star pulsation frequency is slowly increasing over time.

Our analysis of the light curves of the SX Phe variables has allowed us to detect two pulsation frequencies in the variable V54, corresponding to the fundamental and first overtone
radial oscillation modes. We did not detect any other frequencies apart from the dominant frequency for the other two SX Phe variables, and therefore we are unable to identify 
their modes of oscillation.

We provide calibrated $V$ and $I$ photometry in the Johnson-Kron-Cousins photometric system, derived from the analysis of a set of standard stars in the field of the cluster,
and instrumental $r$ photometry, for all of the 41 confirmed variables in NGC~6981 in our field of view. This data is available in electronic form (see Supporting Information) and the format is
shown in Table~\ref{tab:vri_phot}. In order to collect the information required to observe these objects into one place, we calculate celestial coordinates for each variable
(see Table~\ref{tab:astrom}) and provide detailed finding charts in Figure~\ref{fig:finding_charts}. Finding charts for the two RR Lyrae variables V27 and V35 that
are not in our field of view may be found in \citet{dic1972a}.

We performed a Fourier decomposition of the light curves for 21 RR0 and 5 RR1 variables with reliable period estimates and suitable phase coverage, and we report the corresponding
Fourier coefficients in Table~\ref{tab:fourier_coeffs}. The Fourier parameters have been used to estimate the metallicity, absolute magnitude, log-luminosity, and effective
temperature for each RR Lyrae star based on empirical, semi-empirical, and theoretical relations available in the literature. Assuming that the RR Lyrae
stars in NGC~6981 are of the same composition, distance and age, then appropriate averages of the derived properties of the RR Lyrae stars may be calculated and employed as estimates
of these properties for the parent cluster. Applying this method, we derive a metallicity of [Fe/H]$_{\mbox{\scriptsize ZW}} \approx -$1.48$\pm$0.03 on the ZW scale for NGC~6981, and
[Fe/H]$_{\mbox{\scriptsize UVES}} \approx -$1.38$\pm$0.03 on the new \citet{car2009} scale. Similarly, we derive mean true distance moduli of 
$\mu_{0} \approx\,$16.117$\pm$0.047 and 16.111$\pm$0.047~mag for the RR0 and RR1 variables, respectively, and corresponding distances to NGC~6981 of
$\sim$16.73$\pm$0.36~kpc and $\sim$16.68$\pm$0.36~kpc, respectively.

The age of NGC~6981 has been estimated by \citet{dot2010} as 12.75$\pm$0.75~Gyr from analysis of {\it Hubble Space Telescope} ACS observations, and our calibrated $V-I$ CMD
cannot compete with their data in terms of accuracy and lack of systematic errors. Hence we simply demonstrate that our CMD data is consistent with the 
isochrones of the \citet{van2006} stellar evolutionary models when interpolated to the age and metallicity of NGC~6981, and if one chooses to ignore the colour effect of the reddening.
It is likely that the reddening is smaller than we assumed in Section~\ref{sec:distance} or that there is a systematic error in the colour of the theoretical 
stellar evolutionary models that we adopted, or both.

\section*{Acknowledgements}

This research has made use of the SIMBAD database (http://simbad.u-strasbg.fr/simbad/),
operated at CDS, Strasbourg, France, which was essential for performing a complete bibliographical search for NGC~6981.
We would like to thank the support astronomers at IAO, Hanle and CREST (Hosakote), for their efforts in
acquiring the data. Our thanks also goes to ESO librarian Uta Grothkopf for hunting down the 
articles of \citet{ros1953} and \citet{nob1957} which are not available through ADS, and to Vincenzo Forch\`i
who translated them from Italian. We appreciate the highly useful comments from the referee Christine
Clement, and from Carla Cacciari and Don VandenBerg, who all helped in improving the paper.
We acknowledge support from the DST-CONACYT collaboration project and the DGAPA-UNAM grant
through project IN114309 at several stages of the work. Roberto Figuera is grateful to ESO for their hospitality and financial support during
three months while working on this paper. I dedicate this work to my little sister Lucia Mu\~niz Santacoloma.

\section*{Supporting Information}
\label{sec:support_info}

Additional supporting information may be found in the online version of this article.

{\bf Table 4.} Time-series $V$, $r$ and $I$ photometry for all the confirmed variables in our field of view, except V27 and V35 which lie outside of our field of view.

Please note: Wiley-Blackwell are not responsible for the content or functionality of any supporting materials supplied by the authors.
Any queries (other than missing material) should be directed to the corresponding author for the article.

\label{lastpage}

\end{document}